# Numerical Analysis of Temperature and Stress Fields in Mass Concrete Based on Average Forming Temperature Method

Sana Ullah[1†], Wu Peng[1†], Ting Peng[1†], Zujin Fan[2], Tianhao Long[2], Yuan Li[1*]

1. Ministry of Education Key Laboratory of Highway Engineering in Special Areas, Chang'an University, Xian,710064, China

2. Sichuan Chuanjiao Road and Bridge Co. Ltd, Guanghan, Deyang, Sichuan, 618300, China

*Corresponding Author: Yuan Li (liyuan_mm@chd.edu.cn).

†These authors contribute equally.

**ABSTRACT**

Mass concrete plays a crucial role in large-scale projects such as water conservancy hubs and transportation infrastructure. Due to its substantial volume and poor thermal conductivity, the accumulation of hydration heat during the curing process can lead to uneven temperature gradients and stress field distribution, which may cause structural cracking. This phenomenon represents one of the critical challenges in quality control for hydraulic dams, bridge piers and abutments, tunnel linings, and similar engineering structures. To ensure structural safety, it is imperative to calculate temperature variations while optimizing and controlling the temperature stress field. In this paper, a novel method for calculating the zero-stress temperature field is proposed, considering the temperature history and hydration heat release increments at various locations within mass concrete during the curing period, the parameter of average forming temperature field is defined to subsequently solve the temperature stress field. Several typical hydration heat release models were selected to calibrate the computational accuracy of the average forming temperature. Based on simulation results, an optimal model was applied to validate the effectiveness of the proposed method through practical engineering case studies. The impacts of casting temperature, ambient temperature during the curing period, and dimensional thickness on temperature-induced stresses were systematically investigated. Additionally, stress variations at different representative points were compared with the overall mean stress distribution. The results demonstrate that this method can more accurately evaluate temperature-induced stresses caused by seasonal temperature variations. This study provides a more reliable computational basis for ensuring the long-term service safety of mass concrete structures.

## 1. Introduction

Mass concrete is widely used in the engineering field, but the problem of temperature cracking always plagues its quality and durability. After concrete is poured, complex temperature and stress fields will occur internally due to the release of hydration heat of cement, temperature fluctuations in the external environment, and its own thermal expansion and contraction characteristics. When the stress exceeds the tensile strength of concrete, cracks will breed. Among them, the average forming temperature, as a key parameter, has a significant impact on the development of concrete thermal cracks. Especially in the low temperature environment in winter, this effect is more complex. Therefore, it is of great theoretical and practical value to explore the role of the average forming temperature in the simulation of thermal cracks in mass concrete. With the rapid progress of urbanization in our country, infrastructure buildings are changing with each passing day, and the scale and complexity of railway, highway, bridge, subway and other engineering projects are increasing day by day. In this context, mass concrete structures have been widely used in modern infrastructure buildings and have become an indispensable and important building material. Its specific applications cover all kinds of concrete bridges, large dams, concrete baseplates in the basements of high-rise buildings, and large equipment foundation caps. These application scenarios fully demonstrate the excellent advantages of mass concrete in providing strength and stability, and also highlight its irreplaceable key position in large-scale engineering projects.

Temperature cracking is a major concern in mass concrete structures due to the heat generated during cement hydration. Accurately simulating and predicting temperature stresses is crucial for preventing cracking. Several studies have investigated methods to calculate concrete temperature stresses based on average forming temperatures in mass concrete simulations. An et al. (2020) proposed a simplified method for real-time prediction of temperature in mass concrete at early ages. They developed a formula considering the effects of cooling pipes, outside environment, and hydration heat. The method was validated using monitoring data from two large arch dams. The authors found that using the average temperature between the core and surface provided good accuracy in predicting overall temperature development. Evsukoff et al. (2006) developed a data mining approach for modeling adiabatic temperature rise during concrete hydration. They showed that using average temperatures from multiple sensors provided robust predictions of overall temperature development. This average temperature data was suitable as input for stress calculation models in mass concrete simulations. Riding et al. (2006) evaluated different temperature prediction methods for mass concrete members. They compared measured temperatures to predictions from several models, including those using average hydration temperatures. The results showed that methods based on average temperatures could provide reasonable estimates of overall thermal behavior, though they may not capture localized effects as accurately. Schackow et al. (2016) analyzed temperature variations in concrete samples due to cement hydration. They found that using the average temperature across the sample volume provided a good representation of overall hydration heat development. This average temperature approach was suitable for estimating bulk thermal stresses in larger concrete elements. Wang et al. (2018) studied temperature control measures and temperature stress of mass concrete during construction in high-altitude regions. They used 3D finite element analysis to optimize the temperature drop process and control temperature gradients. The researchers demonstrated that using average temperatures across different zones of the structure was effective for simulating overall thermal behavior and stresses. Zhao et al. (2020) investigated the coupled effects of hydration, temperature and humidity in early-age cement-based materials. Their experimental and theoretical analysis demonstrated that an average temperature method could adequately capture the overall thermo-hydration behavior for stress calculations, especially for larger concrete elements where temperature gradients are less pronounced. Mass concrete structures are prone to thermal cracking due to temperature differentials caused by hydration heat. Finite element analysis has been shown to accurately predict temperature distributions and thermal stresses in mass concrete (Anh Kiet Bui & Trong-Chuc Nguyen, 2020; Nguyen-Trong Ho et al., 2020). Factors influencing temperature development include placing temperature, mix design, and environmental conditions (Anh Kiet Bui & Trong-Chuc Nguyen, 2020; M. Tia et al., 2010). Recent research has developed advanced models incorporating hydration, temperature, humidity, and constraint factors to assess temperature damage (Liguo Wang et al., 2024). To mitigate cracking risk, strategies such as using low-temperature-rising polymers, adequate insulation, and appropriate placing temperatures have been proposed (Liguo Wang et al., 2024; M. Tia et al., 2010). While maximum temperature differentials are commonly used to control cracking, supplementary stress analysis is recommended to ensure concrete strength is not exceeded (M. Tia et al., 2010). Isothermal calorimetry testing is suggested for determining heat generation input for finite element modeling (M. Tia et al., 2010).

The hydration heat of concrete refers to the heat released during the hydration process of cement, and it is a key factor affecting the temperature stress of concrete. It is generally believed that the hydration mechanism of cement includes three basic processes: crystal nucleation and crystal growth (NG), phase boundary reaction (I) and diffusion (D)[11]. The hydration heat of cement is correspondingly divided into three stages: induction period, main hydration peak period and later hydration period [12]. In the calculation and research of hydration heat release of concrete, since the amount of hydration heat release in the first stage is extremely small and usually negligible, the hydration heat release model is generally calculated from the second stage [13]. The research on the hydration heat of mass concrete in foreign countries started earlier, and it mostly originated in the field of hydraulic engineering. As early as the 1930s, the US Reclamation Bureau proposed an empirical model of the hydration heat index of mass concrete based on a large amount of experimental data [14], but the model did not consider the effect of temperature and mainly depended on the age of concrete. Since then, Hansen et al. [15] have considered the effect of temperature on the hydration heat release rate, and proposed the equivalent age based on the concept of maturity; James [16] based on the hydration heat release rate proposed a calculation formula for the effect of age and temperature on the hydration rate. Kaszynska (2002) [17] investigated the relationship between the hydration heat of high-performance concrete and the early compressive strength under variable heat conditions, and pointed out that the thermal stress value generated during the hydration exothermic process of cement is crucial for evaluating the early performance of mass concrete during the setting and hardening stage. Poppe et al. (2006) [18] carried out a study on filler-rich binders in self-compacting concrete and modified the existing hydration model of mixed cement systems, emphasizing the importance of cement-to-powder ratio in filler-rich cement-based systems, which provides insight into the hydration process in concrete mixtures of different filler systems. Kim (2010) [19] focused on the heat generated by cement hydration and its impact on mass concrete pouring, especially in the context of using fly ash instead of cement in concrete dams. The study characterized the chemical composition and fineness of cement and fly ash, providing insights into the effects of different materials on the hydration heat of concrete. He et al. (2011) [20] focused on the thermal stress induced by hydration heat of large caps of railway cable-stayed bridges, emphasizing the necessity of spatial distribution studies and temperature control measures during construction. Peng et al. (2016) [21] conducted an experimental study of the hydration heat temperature field of hollow concrete piers. By monitoring the hydration heat temperature distribution of concrete piers, key nodes in the hydration process, such as mold release time, were identified to effectively prevent surface cracking. Ismail et al. (2016) [22] investigated the effect of vinyl acetate wastewater on reducing the hydration heat of concrete. They used vinyl acetate wastewater as a polymer modifier in concrete to control the setting time and heat generation during hydration, providing a new method for managing the hydration heat in concrete mixtures. Mbogo (2016) [23] characterized the microstructure of Portland cement slurry using self-shrinkage and hydration heat parameters, aiming to deepen the understanding of the mechanism of self-shrinkage and hydration heat of cement slurry. Lin et al. [24] developed a finite element model for analyzing the heat-induced stress and associated cracking risks of early concrete members. Fortran subroutines of Abaqus were created to enable the application of material properties in thermal stress calculations. Early elastic modulus, strength development, and tensile and compressive creep behavior were incorporated into the model. Shanhan et al. [25] investigated the effect of concrete temperature rise induced by cement hydration on large concrete structures. Supplementary cementitious materials were used instead of cement to reduce concrete temperature rise. Isothermal measurements were performed on slurries combining supplementary cementitious materials and chemical admixtures. Equations were established based on statistical experimental designs to predict the effect of chemical admixtures and supplementary cementitious materials on total heat. Bažant et al. (2018) [26] investigated the effects of temperature on water diffusion, hydration rate, creep, and shrinkage in concrete structures, focusing on strain and stress modeling at high temperatures, particularly in the context of concrete nuclear power plant structures and high-rise building fires, emphasizing the importance of understanding heat and water volume changes in concrete. Klemczak d et al. (2018) [27] investigated the tensile thermal stress generated in concrete by temperature changes due to heat of hydration to control early cracking of large foundation slabs. Chen et al. (2021) [28] proposed a kinetic model of adiabatic temperature rise in concrete. Factors such as initial temperature and activation energy were considered through thermodynamic analysis and validated in adiabatic temperature rise tests at different initial temperatures. The results showed that the model was superior to traditional models. Liu et al. (2021) [29] established a correlation between the influencing variables and the hydration heat temperature through a support vector machine regression (SVR) model to achieve a short-term prediction of the temperature of mass concrete. The results show that the SVR model can effectively predict the temperature within 2-3 days. Xie et al [30] obtained the parameters of the cement hydration kinetics model suitable for the composite cementing system through the back-propagation (BP) neural networks algorithm, and developed a finite element calculation model based on it to study the time evolution of the three-dimensional temperature field under different heat source functions. The results show that the cement hydration kinetics model is suitable for the calculation of the temperature field of mass concrete as a new heat source function.

The hydration heat generated during the construction of mass concrete and the cracks that may follow it pose serious challenges to the quality of large-scale projects. Due to its large volume, large amount of cementitious materials, and poor thermal conductivity, the heat released by the hydration reaction is difficult to dissipate rapidly during the pouring process, resulting in a sharp rise in internal temperature, while the surface temperature is usually relatively low due to the influence of ambient temperature. This temperature difference can lead to uneven internal expansion and temperature stress, which may lead to cracks, endangering the safety and durability of the structure. In addition, after concrete is formed, when the external temperature is low, it is difficult to synchronize the temperature exchange between the inside and outside of the concrete, and the difference in internal and external shrinkage will also induce temperature stress, which may also lead to cracks. Therefore, special attention must be paid to temperature control and crack prevention during the construction of mass concrete. It is necessary to deeply explore the characteristics and crack formation mechanism of mass concrete structures, actively explore crack control methods, build a simulation model of mass concrete temperature field based on engineering examples, and verify the accuracy of the model through actual data. In order to reduce temperature cracks in mass concrete construction, and comprehensively evaluate the potential and actual effect of new materials and new technologies in mass concrete temperature crack control. This study aims to evaluate various hydration heat models for mass concrete and assess their accuracy against experimental data.

## 2. Theoretical Framework

The theoretical framework is essential for understanding how heat hydration and heat conduction influence the Average Hydration Temperature Method (AHTM) in mass concrete. This section presents a detailed explanation of the concepts, supported by mathematical formulations, figures, and tables extracted from experimental and numerical studies. When cement hydrates, it releases heat, leading to a temperature rise within the concrete. If this heat is not adequately dissipated, it can cause thermal stress, leading to cracking and durability issues. The interaction between heat hydration and heat conduction determines how temperature is distributed within mass concrete structures, influencing crack formation, structural integrity, and long-term performance. The AHTM provides a systematic approach to accurately predicting hydration temperature evolution, integrating heat conduction and stress analysis for better thermal control. This section discusses the governing principles of heat transfer, hydration heat generation, and the role of AHTM in predicting and mitigating thermal stress.

### 2.1 Heat Hydration Process and Heat Conduction in Concrete Structures

Heat hydration is an exothermic reaction that occurs when cement reacts with water, releasing heat and forming hydration products such as calcium silicate hydrate (C-S-H) and calcium hydroxide. This process is influenced by various factors, including cement composition (higher C3S and C3A content increases heat generation), water-to-cement ratio (lower water content results in higher localized temperatures), ambient conditions (temperature and humidity impact the rate of hydration), and admixtures (which influence hydration kinetics and temperature evolution). The heat release rate follows a specific pattern: an initial rapid rise, a dormant phase, followed by a peak reaction phase. Understanding this behavior is crucial for predicting temperature changes and stress formation in mass concrete structures.

The heat generated during hydration is transferred through the concrete via conduction, governed by Fourier's law:

$$q = -\lambda \nabla T$$

where:
q = Heat flux (W/m²)
λ = Thermal conductivity of concrete (W/m°C)
∇T = Temperature gradient (°C/m)

The heat conduction equation, incorporating hydration heat, is:

$$\frac{\partial T}{\partial t} = \frac{\lambda}{c\rho}\left(\frac{\partial^2 T}{\partial x^2} + \frac{\partial^2 T}{\partial y^2} + \frac{\partial^2 T}{\partial z^2}\right) + \frac{Q}{c\rho}$$

where:
c = Specific heat capacity (J/kg°C)
ρ = Density of concrete (kg/m³)
Q = Heat generation rate from hydration (W/m³)

This equation models how heat spreads through concrete, influencing temperature distribution and stress formation. The temperature difference between the concrete core and its surface creates a temperature gradient, leading to thermal stress. If this stress exceeds the tensile strength of concrete, cracks form, compromising structural integrity. Factors affecting thermal stress include cooling rate (faster cooling increases tensile stress), structural dimensions (thicker sections retain more heat), and boundary conditions (heat loss varies based on exposure).

Thermal cracking is a major durability concern, requiring temperature control strategies such as insulation, cooling pipes, staged concrete pouring, and optimized mix designs to minimize temperature gradients and mitigate stress development. Effective implementation of these measures ensures the structural integrity and longevity of mass concrete applications, preventing premature failures due to thermal stresses.

## 2.2 Hydration Exothermic Model

hydration exothermic models explore the heat release process during cement hydration in concrete, integrating multiple mathematical models to describe this complex phenomenon. Traditional hydration models, such as the exponential model, double exponential model, and hyperbolic model, are based primarily on age-dependent hydration kinetics. These models typically take the form:

$$Q(t) = Q_\infty (1 - e^{-at}) \quad (3.1)$$

where Q(t) represents the cumulative heat release at time t, Q∞ is the ultimate heat release, and k is the hydration rate constant. However, these models often fail to capture the effects of microstructure, mineral interactions, and temperature variations.

To enhance accuracy, an equivalent age model based on the Arrhenius equation is introduced to quantify temperature-dependent hydration kinetics:

$$t_e = \int_0^t e^{\frac{E_a}{R}\left(\frac{1}{T_{ref}} - \frac{1}{T}\right)} dt \quad (3.2)$$

where te is the equivalent curing age, Ea is the activation energy, R is the universal gas constant, T is the absolute temperature, and $T_0$ is the reference temperature. This model corrects for variations in temperature, making it more suitable for real-world applications.

Another model incorporated in the study is the degree of hydration model, which defines the degree of hydration α as:

$$\alpha = \frac{Q_0}{Q_{max}} \quad (3.3)$$

where $Q_0$ is the heat released at time ttt and $Q_{max}$ is the total heat potential of the cement. The hydration rate is then given by a reaction kinetics equation:

$$\frac{dQ}{dt} = A e^{\frac{-E_a}{R(T+273.15)}} \cdot \left(1 - \left(\frac{Q_0}{Q_{max}}\right)\right) m \quad (3.4)$$

where A is the pre-exponential factor, and m represents the reaction order. The study determines mmm through experimental fitting, revealing variations based on cement composition and concrete strength grades.

In addition to these formulations, the document discusses adiabatic temperature rise models, which account for heat accumulation and environmental heat exchange:

$$T_{ai} = T_0 + \theta(t) \quad (3.5)$$

where T(t) is the concrete temperature at time t, $T_0$ is the initial temperature, $\theta(t)$ is the concrete density and specific heat capacity.

Experimental data confirm that heat evolution varies significantly with cement type, water-to-cement ratio, and curing conditions. The document emphasizes that these mathematical models are essential for predicting hydration heat, reducing thermal stress, and mitigating crack formation in mass concrete structures. Figures illustrating temperature evolution, hydration degree curves, and reaction kinetics plots provide visual validation of the models, reinforcing their applicability in structural engineering.

Similarly, the composition, physical parameters, heat release, and maximum adiabatic temperature appreciation of several different strength grades of concrete are summarized in **Table 3.4**.

**Table 3.4 Table of concrete parameters**

| strength | $m_0$ | $m_1$ | $m_2$ | $Q_\infty$ | k | $Q_\infty^*$ | c | $\rho$ | $\theta_{max}$ |
|---|---|---|---|---|---|---|---|---|---|
| C40 | 295 | 60 | 40 | 377 | 0.955 | 360 | 960 | 2400 | 61.7 |
| C45 | 280 | 60 | 60 | 377 | 0.92 | 346.8 | 960 | 2410 | 60 |
| C35 | 239 | 55 | 73 | 377 | 0.885 | 333.6 | 960 | 2390 | 53.4 |
| C30 | 208 | 48 | 64 | 377 | 0.885 | 333.6 | 960 | 2385 | 46.6 |

Using a Python program, taking time t (set to 28d) as the abscissa, the data of the formula is compared through the fitting process. The optimal value of the parameter 'm' in, the program is as follow. The fitting curves of different strength grades of concrete are shown in **Figure 2.4**;

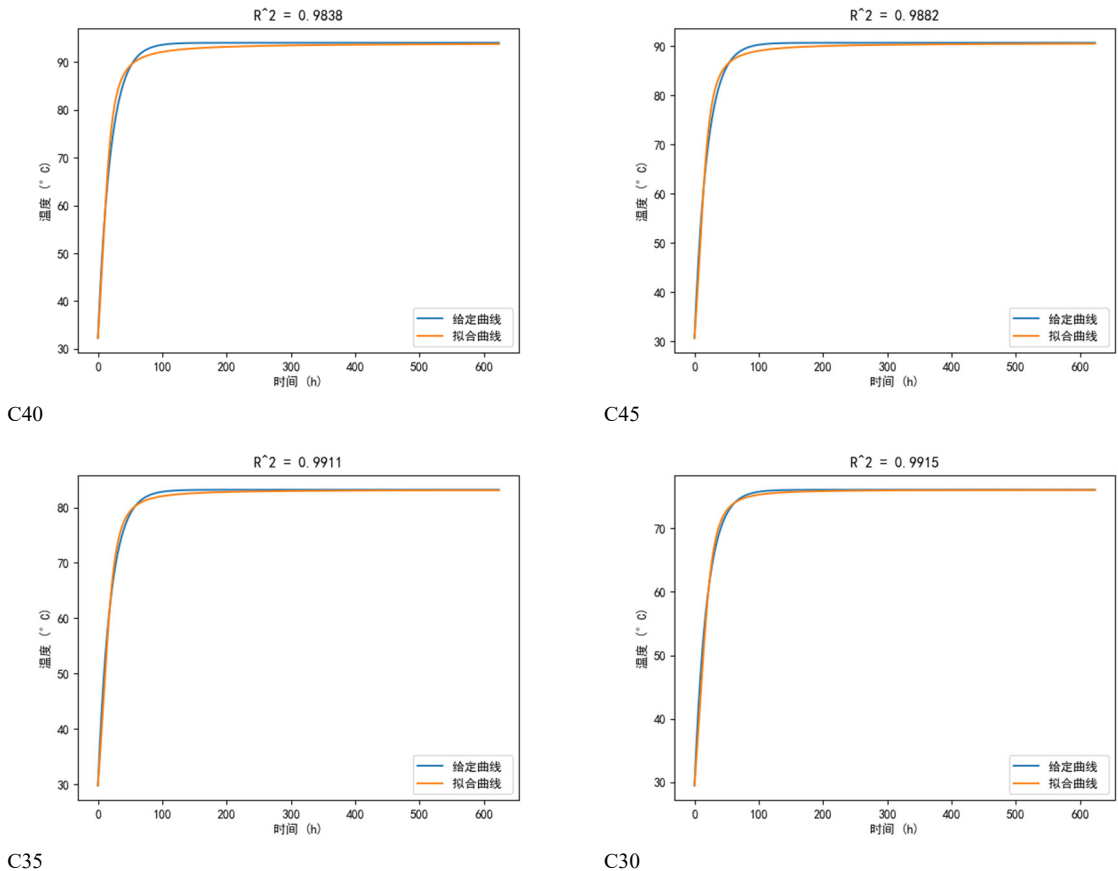

C40  C45

C35  C30

**Figure 2.4: Fitting curve**

# 3. Materials and Method
## （1） Simulation and Verification of Hydration Heat Release Model
### a) Finite Element Simulation

The finite element method (FEM) is a numerical technique used to solve complex engineering and physical problems. Its basic principle is to divide the continuum into small, finite elements, and the solution in each element is approximated by a simple function. By establishing local equations for each element and synthesizing them into a global system of equations, FEM transforms complex partial differential equations into a system of linear equations. Finally, by solving this system of equations, an approximate solution to the system is obtained.

COMSOL Multiphysics is a numerical simulation software package that uses the finite element method to solve various scientific research and engineering problems described by partial differential equations (PDEs). Based on the PDE model, the software can easily define and solve multi-physics coupling problems. It integrates pre-processing, solving and post-processing functions and is a comprehensive finite element calculation and analysis (CAE) tool.

The simulation of the hydration heat release model uses the solid heat transfer module of COMSOL, which mainly solves the heat transfer problem based on the heat conduction equation (heat diffusion equation). This equation is the basic equation that describes the heat transfer in solids. For transient studies, the equation is in the following form:

$$\rho c \frac{\partial T}{\partial t} = \lambda \nabla^2 T + Q \qquad (4.1)$$

$\lambda$ —thermal conductivity, $W/(m \cdot ℃)$。

$c$ —concrete specific heat capacity, $J/(kg \cdot ℃)$；

$\rho$ —Concrete density, $kg/m^3$；

Q—Heat released per unit volume per unit time, $J/(m^3 \cdot h)$；

$\nabla^2 T$ —Describes operator for temperature and the spatial diffusion of heat.

In view of the complexity of engineering conditions and the variability of environmental conditions, the temperature field and stress field of mass concrete blocks are affected by a variety of factors. In order to simplify the numerical simulation of the structural model, the following assumptions are made:

(1) Assume that the concrete material is uniform and isotropic in physical properties.
(2) It is assumed that during the concrete pouring process, the initial mold entry temperature throughout the structure remains the same, and there is no temperature difference; at the same time, the surrounding environment of the concrete structure also has the same temperature conditions.
(3) It is assumed that when the concrete block is heat exchanged with the outside, the heat release coefficient of each side is the same.

## （2） Simulation of Concrete Temperature Field:
### a) Simulation Specific Process

The temperature field caused by the hydration heat of mass concrete is simulated by COMSOL multi-physics finite element method. The specific process is as follows:

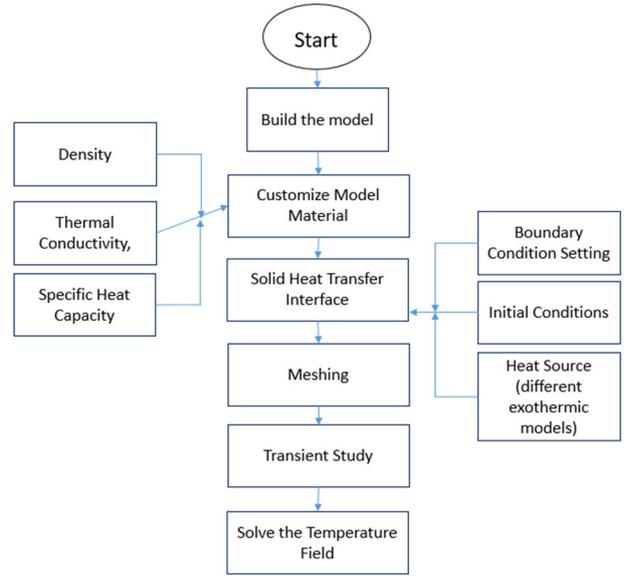

**Figure 3 1 Flow chart of temperature field solution**

### b) Concrete Temperature Field Simulation Model

This simulation uses a three-dimensional model. The geometric dimensions of the model are based on the 1m×1m×1m large-volume concrete block in the temperature rise test of different concrete strength grades under the same reinforcement ratio in the literature [26]. In order to ensure the consistency between the simulation and the actual situation, the established geometric model should be consistent with the concrete block equipped with steel bars in the test, as **Figure 2.5** below;

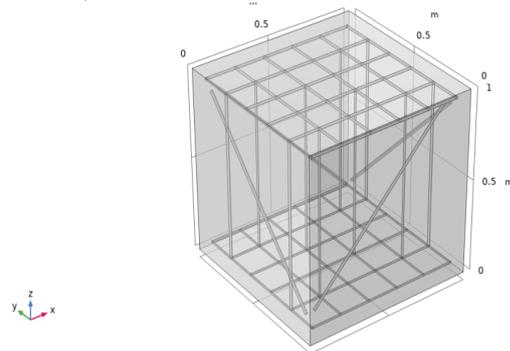

**Figure 2.5: Schematic diagram of mass concrete dimensions**

At this point, the established geometric model has not yet been converted into a finite element model, and it needs to be meshed to convert it into a finite element model, as shown in **Figure 2.6** below;

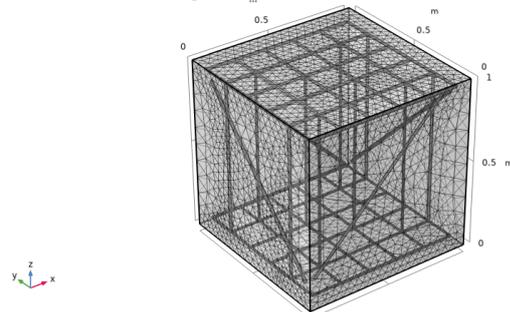

**Figure 2.6: model mesh division**

The free triangle mesh method was used for mesh generation, and the normalized mesh was selected. The generation process was automatically controlled by the selected physical field. The mesh data after division is shown in **Table 3.7**.

**Table 3.7: Grid statistics**

| Grid Parameters | quantity |
|---|---|
| Mesh Vertices | 20858 |
| Tetrahedral element | 117212 |
| Triangular unit | 16694 |
| Number of edge elements | 7365 |

| Number of vertex units | 760 |
|---|---|
| Minimum unit mass | 0.1192 |
| Average unit quality | 0.5812 |
| Unit volume ratio | 4.131E-7 |

### c) Model Parameter Selection
(1) Concrete thermal conductivity
The thermal conductivity of concrete is an important thermal performance parameter in the analysis of hydration heat conduction of concrete, which reflects the heat conductivity of concrete materials. Model parameter results shown in **Table 3.9.** Schindler established a thermal conductivity function based on hydration degree

$$\lambda(\alpha) = \lambda_u(1.33 - 0.33\alpha) \quad (1.1)$$

$\lambda_u$ —final thermal conductivity $W/(m \cdot ℃)$, here its 2.325

According to whether the hydration degree factor is considered in various hydration exothermic models, the value of the thermal conductivity of different hydration exothermic models is different, as shown in **Table 3.8**

**Table 3.8 Valuation methods of thermal conductivity of different exothermic models**

| exothermic model | thermal conductivity / $W/(m \cdot ℃)$ |
|---|---|
| I | 2.325 |
| II | 2.325[1.33-0.33α(t_e)] |
| III | 2.325[1.33-0.33α(t)] |
| IV | 2.325(1.33-0.33Q_0/Q_{max}) |

(2) Concrete heat release coefficient
The heat transfer coefficient of concrete refers to the efficiency of heat transfer between concrete and the outside world. The heat transfer coefficient of exposed concrete surface is related to the wind speed. If there is a thermal insulation layer on the surface, the surface heat transfer coefficient mainly depends on the thickness and thermal conductivity of the thermal insulation layer. In the literature, the surface of the mass concrete test block is covered with a film, and the film is covered with cotton felt for curing. Considering the material of the thermal insulation layer and its thickness, the surface heat transfer coefficient β of concrete can be obtained by calculating $22 W/(m^2 \cdot ℃)$.

(3) Reinforcement thermal parameters

**Table 3-6 Thermal parameters of steel reinforcement**

| parameter name | Value |
|---|---|
| Density (kg/m ') | 78502 |
| Thermal conductivity (W/(m- "O")) | 45e |
| Specific heat capacity (J/(Kg. C) | 500 |

(4) Ambient temperature
In order to more realistically reflect the influence of temperature fluctuations on coagulation temperature changes, the cosine function is used to obtain the periodic function of local air temperature transformation. The formula is as follows:

$$T_a = 28.7 + 32.12\cos[\frac{\pi}{6}(t-15)] \quad (1.2)$$

**Table 3.9: Model Parameter Results**

| Object | parameter | numerical value | unit |
|---|---|---|---|
| Concrete | Thermal conductivity | 2.325[1.33-0.33α(t)] | $W/(m \cdot ℃)$ |
| | Heat release coefficient | 22 | $W/(m^2 \cdot ℃)$ |
| | Initial temperature | 30 | ℃ |
| Steel bar | Thermal conductivity | 45 | $W/(m \cdot ℃)$ |
| | Density | 7850 | $kg/m^3$ |

### （3） Temperature Field Simulation Results
In order to study the difference in temperature rise of mass concrete blocks with different strength grades under the same reinforcement ratio in the reference, the temperature changes of 5 measuring points in the concrete block structure are monitored and recorded, as shown in **Figure 2.7**. This paper will extract the highest and lowest temperatures and corresponding times of each monitoring point under different concrete strength models under different exothermic models from the simulation analysis results, and compare them with the actual data to verify and evaluate the temperature rise effect of various exothermic models.

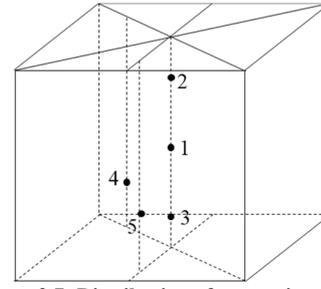

**Figure 2.7: Distribution of measuring points**

After obtaining the temperature field simulation results of the traditional hydration heat model (I), the equivalent age hydration exothermic model (II), the adiabatic heating derivation model (III) and the multi-factor hydration heat model (IV), this paper extracts the temperature maxima and minima of 5 measuring points under four concrete strength grades under different exothermic models and their corresponding times. The data of the hydration exothermic model I~ IV are sorted into **Tables 3.2 to 3.10** respectively.

**Table 3.3 Exothermic Model I Simulated Temperature Maximum and Time**

| measuring point | Concrete strength grade | | | | | | | |
|---|---|---|---|---|---|---|---|---|
| | C40 | | C45 | | C35 | | C30 | |
| | Temp/°C | Time/h | Temp/°C | Time/h | Temp/°C | Time/h | Temp/°C | Time/h |
| 1 | 60.7 | 25.3 | 58.9 | 25.4 | 54.7 | 25.2 | 47.4 | 21.3 |
| 2 | 53.2 | 25.6 | 51.8 | 25.0 | 48.7 | 24.9 | 39.1 | 21.5 |
| 3 | 50.6 | 24.8 | 49.2 | 25.3 | 46.0 | 25.3 | 41.3 | 20.6 |
| 4 | 54.0 | 25.1 | 52.5 | 25.2 | 49.1 | 25.5 | 42.0 | 21.9 |
| 5 | 55.8 | 24.9 | 54.2 | 25.2 | 50.6 | 25.8 | 43.9 | 21.2 |

**Table 3.4 Exothermic model i Simulated temperature minimum and time**

| measuring point | Concrete strength grade | | | | | | | |
|---|---|---|---|---|---|---|---|---|
| | C40 | | C45 | | C35 | | C30 | |
| | Temp/°C | Time/h | Temp/°C | Time/h | Temp/°C | Time/h | Temp/°C | Time/h |
| 1 | 25.9 | 182.3 | 22.0 | 199.8 | 26.0 | 159.1 | 23.3 | 132.2 |
| 2 | 29.3 | 112.4 | 29.2 | 105.3 | 28.6 | 101.7 | 23.0 | 104.8 |
| 3 | 23.9 | 181.5 | 21.4 | 200.1 | 24.0 | 160.2 | 22.3 | 131.1 |
| 4 | 25.1 | 182.3 | 21.3 | 199.2 | 25.8 | 129.3 | 23.5 | 105.3 |
| 5 | 25.1 | 181.0 | 21.6 | 199.5 | 26.2 | 129.5 | 22.8 | 132.1 |

**Table 3.5 Exothermic model ii Simulated temperature maximum and time**

| measuring point | Concrete strength grade | | | | | | | |
|---|---|---|---|---|---|---|---|---|
| | C40 | | C45 | | C35 | | C30 | |
| | Temp/°C | Time/h | Temp/°C | Time/h | Temp/°C | Time/h | Temp/°C | Time/h |
| 1 | 72.6 | 23.3 | 69.4 | 23.2 | 62.7 | 24.1 | 56.8 | 25.1 |
| 2 | 62.1 | 23.4 | 59.9 | 24.1 | 54.8 | 25.2 | 50.2 | 24.7 |
| 3 | 59.4 | 21.8 | 57.0 | 22.3 | 51.9 | 23.8 | 47.4 | 24.3 |
| 4 | 63.0 | 22.1 | 60.6 | 24.3 | 55.2 | 25.0 | 50.3 | 25.8 |
| 5 | 65.8 | 22.2 | 63.1 | 23.8 | 57.3 | 24.2 | 52.1 | 25.4 |

**Table 3.6 Exothermic model ii Simulated temperature minimum and time**

| measuring point | Concrete strength grade | | | | | | | |
|---|---|---|---|---|---|---|---|---|
| | C40 | | C45 | | C35 | | C30 | |
| | Temp/°C | Time/h | Temp/°C | Time/h | Temp/°C | Time/h | Temp/°C | Time/h |
| 1 | 26.1 | 151.3 | 25.8 | 151.5 | 25.8 | 151.3 | 24.8 | 172.2 |
| 2 | 27.4 | 124.8 | 27.0 | 125.2 | 26.8 | 125.5 | 27.8 | 110.1 |
| 3 | 24.1 | 150.3 | 23.9 | 151.0 | 23.8 | 151.1 | 23.2 | 173.2 |
| 4 | 24.9 | 151.1 | 24.8 | 151.2 | 24.7 | 151.3 | 25.5 | 133.4 |
| 5 | 25.1 | 151.8 | 24.9 | 150.9 | 24.8 | 151.2 | 25.7 | 134.5 |

**Table 3.7 Exothermic Model iii Simulated Temperature Maximum and Time**

| measuring point | Concrete strength grade | | | | | | | |
|---|---|---|---|---|---|---|---|---|
| | C40 | | C45 | | C35 | | C30 | |
| | Temp/°C | Time/h | Temp/°C | Time/h | Temp/°C | Time/h | Temp/°C | Time/h |
| 1 | 57.2 | 16.4 | 55.7 | 16.5 | 52.3 | 17.8 | 49.3 | 18.0 |
| 2 | 47.4 | 20.1 | 46.8 | 21.2 | 44.8 | 20.9 | 43.0 | 22.1 |
| 3 | 46.4 | 15.5 | 45.5 | 15.9 | 43.1 | 16.8 | 41.2 | 16.8 |
| 4 | 48.6 | 13.8 | 47.7 | 14.1 | 45.2 | 15.9 | 43.0 | 15.9 |
| 5 | 50.9 | 15.8 | 49.9 | 15.8 | 47.2 | 16.3 | 44.8 | 16.3 |

**Table 3.8 Exothermic model iii Simulated temperature minimum and time**

| measuring point | C40 Temp /°C | C40 Time /h | C45 Temp /°C | C45 Time /h | C35 Temp /°C | C35 Time /h | C30 Temp /°C | C30 Time /h |
|---|---|---|---|---|---|---|---|---|
| 1 | 26.8 | 121.3 | 24.9 | 127.3 | 24.8 | 127.1 | 24.7 | 126.3 |
| 2 | 26.7 | 111.6 | 23.6 | 126.8 | 23.6 | 126.5 | 23.6 | 126.0 |
| 3 | 24.6 | 121.8 | 23.4 | 127.5 | 23.3 | 127.4 | 23.2 | 126.2 |
| 4 | 25.8 | 111.9 | 23.7 | 126.4 | 23.6 | 126.4 | 23.5 | 126.2 |
| 5 | 25.9 | 121.5 | 24.0 | 127.1 | 23.9 | 126.3 | 23.8 | 126.7 |

**Table 3.9 Exothermic Model IV Simulated Temperature Maximum and Time**

| measuring point | C40 Temp /°C | C40 Time /h | C45 Temp /°C | C45 Time /h | C35 Temp /°C | C35 Time /h | C30 Temp /°C | C30 Time /h |
|---|---|---|---|---|---|---|---|---|
| 1 | 62.1 | 21.6 | 57.6 | 23.4 | 53.5 | 25.6 | 54.5 | 25.2 |
| 2 | 55.0 | 20.3 | 54.3 | 21.8 | 50.6 | 21.8 | 51.1 | 21.3 |
| 3 | 50.3 | 20.8 | 48.4 | 23.0 | 43.4 | 24.8 | 45.9 | 24.8 |
| 4 | 53.8 | 24.6 | 51.3 | 26.1 | 46.3 | 27.6 | 49.0 | 27.3 |
| 5 | 54.8 | 22.6 | 52.9 | 25.5 | 46.2 | 27.3 | 50.4 | 26.5 |

**Table 3-10 Exothermic Model IV Simulated Temperature Minimum and Time**

| measuring point | C40 Temp /°C | C40 Time /h | C45 Temp /°C | C45 Time /h | C35 Temp /°C | C35 Time /h | C30 Temp /°C | C30 Time /h |
|---|---|---|---|---|---|---|---|---|
| 1 | 27.0 | 148.2 | 28.5 | 132.5 | 28.6 | 132.5 | 28.5 | 123.2 |
| 2 | 27.4 | 125.6 | 27.8 | 131.1 | 28.1 | 131.9 | 28.4 | 122.6 |
| 3 | 24.9 | 148.6 | 25.9 | 130.9 | 26.0 | 132.4 | 25.9 | 125.5 |
| 4 | 26.2 | 147.3 | 27.2 | 131.6 | 27.2 | 133.4 | 27.3 | 124.6 |
| 5 | 26.2 | 147.1 | 27.3 | 132.6 | 27.4 | 132.1 | 27.4 | 123.8 |

(4) **Verification and evaluation of exothermic model**

The maximum and minimum temperature values of different measuring points for the measured strength of four kinds of concrete in the literature and their corresponding times are shown in Table 3 11 and Table 3 12.

**Table 3 11 Measured table of maximum temperature and time**

| measuring point | C40 Temp /°C | C40 Time /h | C45 Temp /°C | C45 Time /h | C35 Temp /°C | C35 Time /h | C30 Temp /°C | C30 Time /h |
|---|---|---|---|---|---|---|---|---|
| 1 | 65.9 | 25.2 | 61 | 25.4 | 59.1 | 26.2 | 52.2 | 25.5 |
| 2 | 65.7 | 17.7 | 64 | 13.8 | 61.5 | 18 | 53.7 | 18.9 |
| 3 | 50.7 | 23.2 | 48.6 | 29.6 | 46.2 | 29.4 | 43.5 | 26.5 |
| 4 | 55.6 | 22.5 | 52.7 | 25.1 | 50.6 | 24.5 | 46.7 | 23.2 |
| 5 | 56.1 | 23.1 | 51.6 | 24.8 | 52.2 | 25.2 | 47.8 | 23.8 |

**Table 3 12 Temperature minimum and time measurement table**

| measuring point | C40 Temp /°C | C40 Time /h | C45 Temp /°C | C45 Time /h | C35 Temp /°C | C35 Time /h | C30 Temp /°C | C30 Time /h |
|---|---|---|---|---|---|---|---|---|
| 1 | 26.4 | 139.3 | 28.3 | 142.3 | 31.9 | 138.2 | 29.6 | 138.2 |
| 2 | 24.4 | 131 | 25.2 | 130.7 | 32.1 | 130.7 | 28 | 134 |
| 3 | 26.3 | 140.2 | 28.9 | 142.3 | 30.3 | 138.2 | 29.1 | 137.3 |
| 4 | 25.1 | 134.3 | 28.9 | 137.3 | 30.1 | 134.8 | 28 | 134 |
| 5 | 25.9 | 136.8 | 28.5 | 135.7 | 30.9 | 134.8 | 29.3 | 134 |

## 4. Calculation of Temperature Stress and Cracking Risk
### 1. Prediction method for cracking area of mass concrete

The concrete forming period is set as 28 days, which is divided into (n) time steps, and the hydration heat release increment per unit volume of concrete and the corresponding temperature are recorded in each time step. The product of the two accumulates to obtain the total heat accumulation effect inside the concrete during the forming period. Divide the total hydration heat release to obtain the average forming temperature, which accurately reflects the average temperature state of the concrete during the entire forming period.

$$T_{avg} = \frac{\sum_{i=1}^{n} \Delta Q_i \cdot T_i}{\sum_{i=1}^{n} \Delta Q_i} \quad (4.1)$$

It is assumed that the concrete does not expand, contract or strain at the average forming temperature, which is used as the boundary point of thermal expansion and contraction of the material. This assumption is based on the understanding of the properties of concrete materials, aims to simplify the complex physical process, and focuses on the temperature stress caused by the external temperature fluctuation after concrete molding. Taking the average forming temperature as the initial state, it is assumed that the internal concrete reaches a relatively stable thermal stress equilibrium state, simplifying the complex initial stress state and facilitating analysis.

$$\varepsilon_T = \alpha_c \cdot (T - T_r) \quad (4.2)$$

The strain generated by the temperature difference is divided into constrained strain $\varepsilon_1$ and actual strain, namely $\varepsilon_2$:

$$\varepsilon_T = \varepsilon_1 + \varepsilon_2 \quad (4.3)$$

The strain generated by the temperature difference is divided into constrained strain $\varepsilon_1$ and actual strain, namely $\varepsilon_2$:

$$\varepsilon_T = \varepsilon_1 + \varepsilon_2 \quad (4.4)$$

The magnitude of the temperature stress generated by the deformation constraint is:

$$\sigma_T = E \cdot \varepsilon_1 \quad (4.5)$$

Where E is the elastic modulus of the concrete.
These assumptions a more accurate prediction of the internal stresses, considering the complex thermal behavior of the concrete during the entire curing phase.
Derivation of Average Molding Temperature
A slender concrete strip is divided into n segments of concrete micro-elements. The n segments of concrete micro-elements are in close contact and arranged. The forming temperatures of concrete micro-elements 1, 2,..., n are T1, T2,..., Tn, respectively, as shown in the figure below :

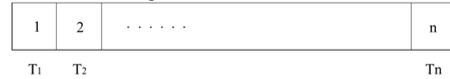

**Figure 4-1 Slender concrete strip**

Each micro element is regarded as a point, and there is currently an ambient temperature Ta. According to the principle of thermal expansion and contraction, the linear strain of the material when the temperature changes is proportional to the temperature change. Let the linear expansion coefficient of concrete be, and the elastic modulus be E. For any point i, the forming temperature is recorded as Ti, and the free expansion or contraction strain at any point is $\varepsilon_i = \alpha_c(T_i - T_a)$:

Since the close contact of n micro points cannot be strained, there will be an interaction force between them, resulting in stress. According to Hooke's law $\sigma = E\varepsilon$, Point I will be subject to compressive stress tensile stress $\sigma_i = E\varepsilon_i$。 In order to keep the entire slender concrete strip in a state of equilibrium, the combined temperature stress of the entire system is 0，that is $\sigma_1 + \sigma_2 + \ldots + \sigma_n = 0$, This leads to :

$$\sum_{i=1}^{n} \sigma_i = 0 \quad (4.6)$$

After substituting for Hooke's law formula :

$$\sum_{i=1}^{n} E\varepsilon_i = 0 \quad (4.7)$$

Expand the strain into a formula for thermal expansion :

$$\sum_{i=1}^{n} E\alpha_c(T_i - T_a) = 0 \quad (4.8)$$

Available after sorting :

$$E\alpha_c(\sum_{i=1}^{n}T_i - nT_a) = 0 \qquad (4.9)$$

Finally, it is deduced：

$$T_a = \frac{\sum_{i=1}^{n}T_i}{n} \qquad (4.10)$$

From the final derivation results, it can be seen that in order to make the temperature stress of the whole system 0, the ambient temperature Ta is equal to the arithmetic average of the molding temperature of all n micro points, and the ambient temperature at this time is defined as the average molding temperature, that is：

$$T_{ave} = T_a = \frac{\sum_{i=1}^{n}T_i}{n} \qquad (4.11)$$

Similar concepts are applied in 2D and 3D scenario's, where the temperature distribution is more complex due to multiple spatial dimensions. The average molding temperature in these cases is still the arithmic mean of temperature at the micro points in the systems mentation of new methods in Finite Element Simulation.

## 2. Implementation of new methods in Finite element

In finite element method (FEM) Is employed to solve the temperature stress field. The process includes calculating the temperature field from the previous section, followed by the using average molding temperature as a strain reference in the simulation to predict stress distribution and cracking risks. The general process is as follows：

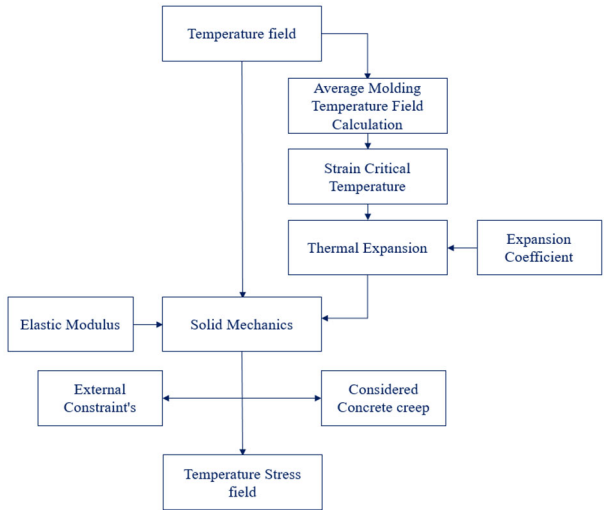

**Figure 4.4: Flowchart of Stress Solution**

### a) Physics Field Selection:

This temperature stress simulation was carried out under the "Thermal-Solid Coupling" interface, and the two physical field modules "Solid Heat Transfer" and "Solid Mechanics" were connected through the multi-physics interface of thermal expansion. The "Solid Heat Transfer" module was already used in the previous chapter, and this time the "Solid Mechanics" physical field needs to be added.

The Solid Mechanics module is mainly used to simulate the mechanical behavior of structures. The system of equations built into this module is usually based on the classical principles of continuum mechanics. The relationship between stress and strain is described by the constitutive equation of the material. For linear elastic materials, i.e. assuming that the material deforms within the elastic range, the relationship between stress and strain can be described by Hooke's law：

$$\sigma = [C] \cdot \varepsilon \qquad (4.12)$$

Where [C] is the stiffness matrix, and $\epsilon$\epsilon$ is the strain tensor.

### b) Calculate Average Molding Temperature Field

In the finite element simulation, according to the model with the best precision performance in the hydration exothermic model in the previous chapter, the heat source function of the mass concrete structure is set, and the temperature distribution at intervals of 0.1h within 28d of the concrete forming period is obtained, as well as the heat increment at a single time interval. Since the heat source function is a power function, the calculation formula of the average forming temperature can be converted to：

$$T_{avg} = \frac{\sum_{i=1}^{n}P_i \cdot \Delta t \cdot T_i}{\sum_{i=1}^{n}P_i \cdot \Delta t} \qquad (4.13)$$

The new method is demonstrated through an engineering example involving the west extension of Chuangye Avenue in Mianyang City, Sichuan, specifically focusing on the cracking risk in the concrete sidewalls of an underpass tunnel. The tunnel's concrete, made of C40 waterproof reinforced concrete, experiences significant temperature fluctuations between summer and winter, leading to cracking. To analyze this, the average molding temperature concept is applied. The modeling process begins by creating geometric models for the sidewalls with varying longitudinal dimensions: 5m, 10m, 15m, and 20m (Figure 4-6). These models are meshed, and the finite element method (FEM) is used to simulate temperature stress. The boundary conditions include sliding constraints on both sides, ensuring vertical deformation is restricted. Table 4.1 provides the tensile strength values for concrete, crucial for comparing stress levels during simulation. The are based on environmental conditions, with the temperature distribution initial and ambient temperatures over a 28-day forming period being calculated to derive the average molding temperature field (Figures 4-8). This field is then used to calculate the temperature stress distribution (Figure 4-9 to Figlure 4-12) under low-temperature winter conditions. The results reveal that when considering the average molding temperature, the internal stresses are higher compared to the standard method that does not factor in the temperature history. The maximum tensile stress is significantly increased, with **Table 4.2** showing the mean tensile stress values, which confirm the increased risk of cracking in the internal regions of the structure. This comparison is further analyzed in **Figures 4-13 to 4-16**, highlighting areas where the tensile stress exceeds the tensile strength of concrete, indicating potential cracking. The study provides a clear step-by-step procedure to model and predict cracking risk based on the average molding temperature, illustrating how the method can be applied to real-world engineering problems.

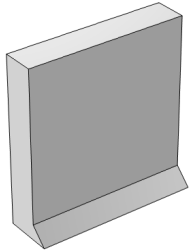

(a)5m

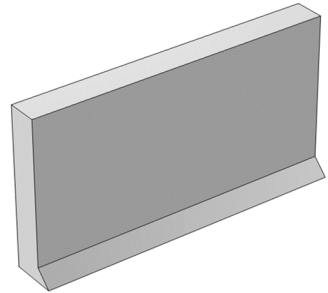

(b)10m

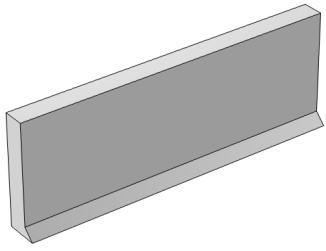
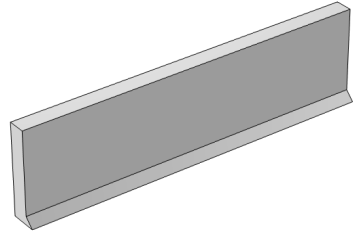

(c)15m          (d)20m

**Figure 4-6 Concrete side wall models of different sizes**

Each size model is meshed and converted into a finite element model, as shown in Figures 4.7：

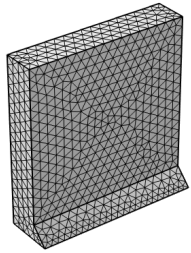
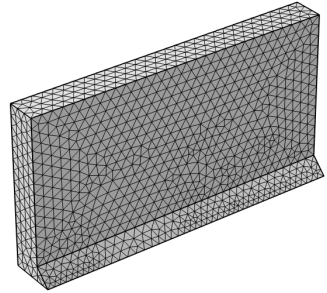

(a)5m          (b)10m

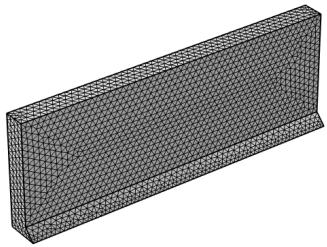
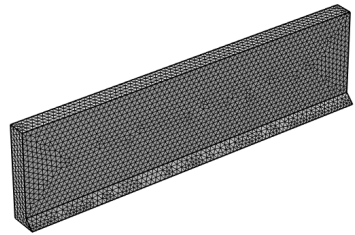

(c)15m          (d)20m

**Figure 4-7 Schematic division of side walls of different sizes**

c) ***Stress Field***

The solution of this study is divided into two processes. First, the average forming temperature field of the concrete sidewall during the whole pouring and forming period is calculated based on the physical field of "solid heat transfer", and then the sidewall is placed under the cold temperature conditions in winter to solve the structural temperature stress distribution map.

（1）*Calculation of average forming temperature field*

On the basis of the solid heat transfer field, a domain ordinary differential equation interface is connected, and the average forming temperature field of the sidewall during the hydration forming period of 28 days is calculated by the formula, as shown in Figure 4.8

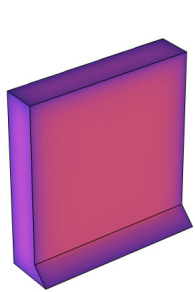
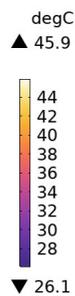
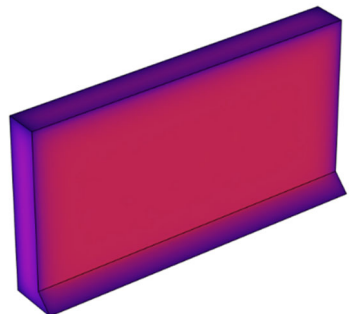
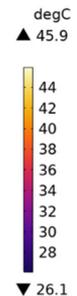

(a)5m          (b)10m

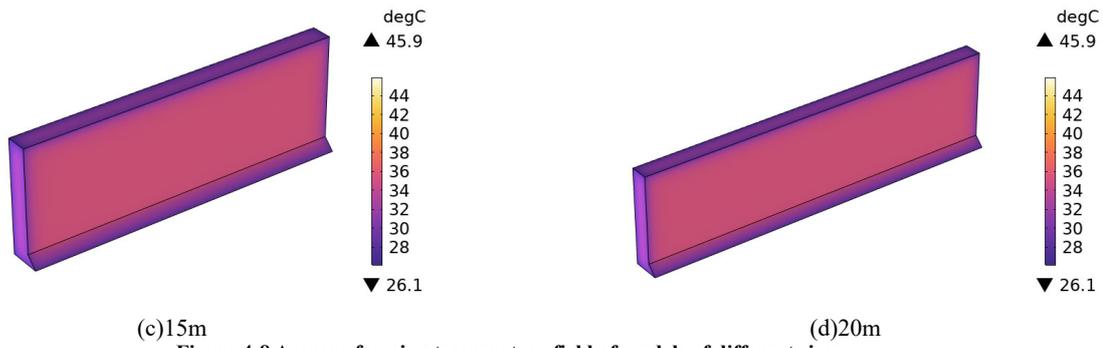

(c)15m          (d)20m

**Figure 4-8 Average forming temperature field of models of different sizes**

It can be seen that the range of the average molding temperature field is consistent under each longitudinal dimension. This is mainly because the external ambient temperature conditions are consistent, providing a unified temperature reference condition for the molding process, so that the average molding temperature field does not change in the range due to different longitudinal dimensions。

*(2) stress field solution*

The average forming temperature field is used at the strain reference temperature as the zero-strain temperature field of the model. Under this temperature field, the structure will not be strained or stressed, that is, the zero-stress temperature field. Then, the physical field of "solid mechanics" is connected, and the model is placed in the cold conditions of 5 ° C in winter for simulation. According to the principle of thermal expansion and contraction, the concrete structure will experience contraction strain at low temperature, and internal stress will be generated due to uneven or constrained strain. For subsequent analysis, a comparison item is added in the research step to calculate the maximum tensile stress distribution without considering the average forming temperature, that is, the initial temperature is set to the zero-stress temperature field. Other conditions are consistent.

The first principal stress distribution diagram of concrete side wall structure of each longitudinal dimension is obtained by finite element method considering the average forming temperature. When the first principal stress is positive, it represents the maximum tensile stress at this point. The maximum tensile stress distribution of concrete side wall structure with different longitudinal dimensions is obtained by finite element method

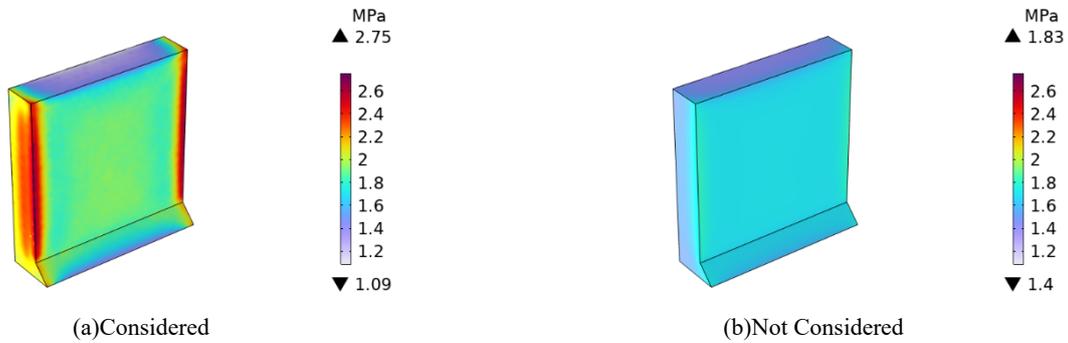

(a)Considered          (b)Not Considered

**Figure 4-9 Longitudinal dimension 5m: Comparison of maximum tensile stress distribution**

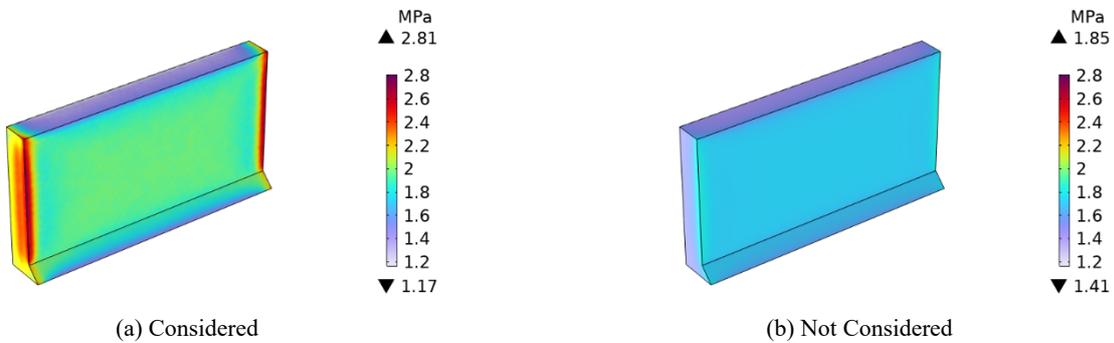

(a) Considered          (b) Not Considered

**Figure 4-10 Longitudinal dimension 10m: Comparison of maximum tensile stress distribution**

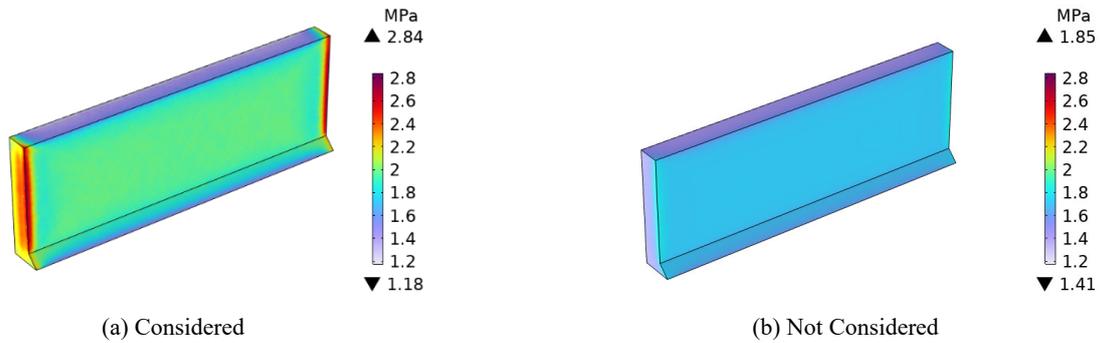

(a) Considered          (b) Not Considered

**Figure 4-11 Longitudinal dimension 15m: Comparison of maximum tensile stress distribution**

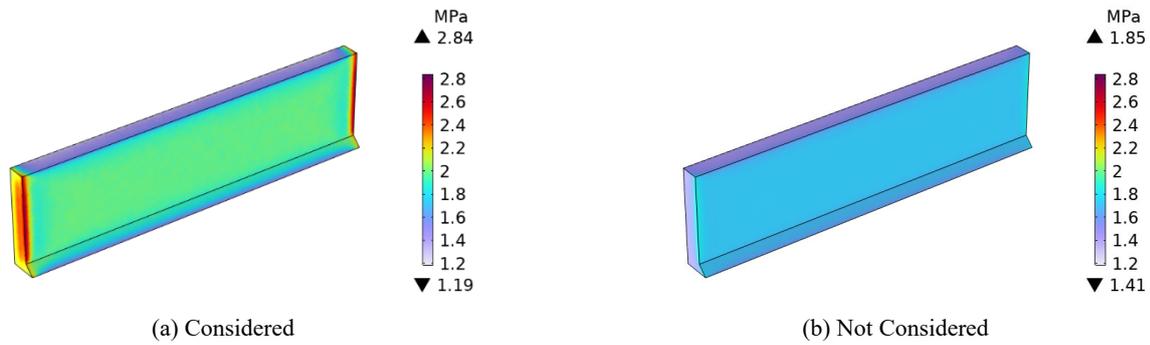

(a) Considered    (b) Not Considered

**Figure 4-12 Longitudinal dimension 20m: Comparison of maximum tensile stress distribution**

### i. Analysis Comparison

As can be seen from Figures 4 9 to 4 12, regardless of whether the average forming temperature is taken into account, the larger stress values are mainly concentrated in the constrained areas on both sides of the structure. This is due to the phenomenon of stress concentration caused by the constraints on the strain of concrete. When the longitudinal dimensions are the same, the maximum value of the first principal stress distribution when the average forming temperature is taken into account is nearly 1 MPa more than when this factor is not taken into account. In view of the characteristics of low tensile properties of concrete materials, the impact of this difference should not be underestimated, and it is very likely to become one of the key factors that lead to increased risk of concrete cracking. Further comparison found that with the increase of longitudinal dimensions, the stress range of the two cases hardly changed.

The maximum tensile stress theory in mechanics of materials is used to judge the possible cracking area of the mass concrete sidewall structure. The standard value of concrete tensile strength can be obtained in the following table :

**Table 4.1 Standard values of concrete tensile strength**

| Intensity grade | C25 | C30 | C35 | C40 | C45 |
|---|---|---|---|---|---|
| $f_{tk}$ (Mpa) | 1.78 | 2.01 | 2.20 | 2.39 | 2.64 |

C40 concrete is used in this project, so the standard value of tensile strength is 2.39MPa. The area greater than the tensile strength in both cases is shown in the figure below :

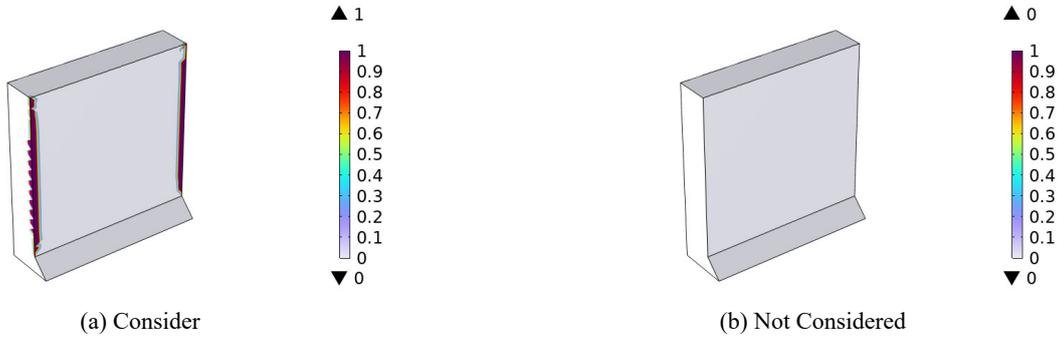

(a) Consider    (b) Not Considered

**Figure 4-13 Longitudinal dimension 5m: distribution of excess tensile strength**

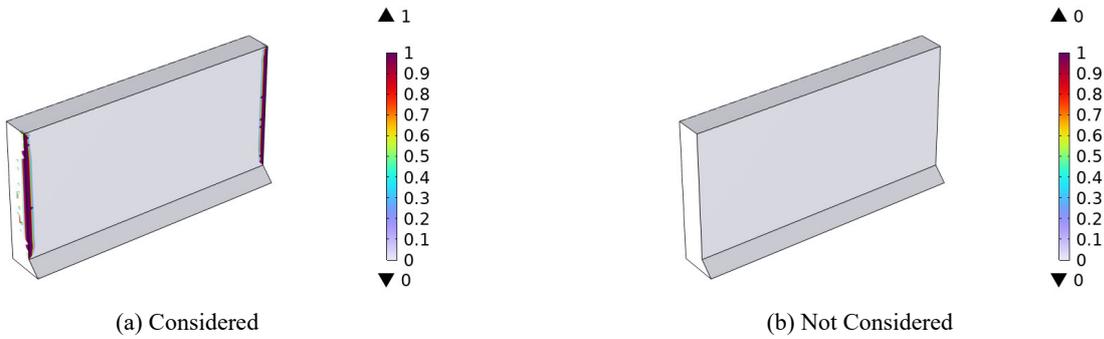

(a) Considered    (b) Not Considered

**Figure 4-14 Longitudinal dimension 10m: distribution of tensile strength exceeding**

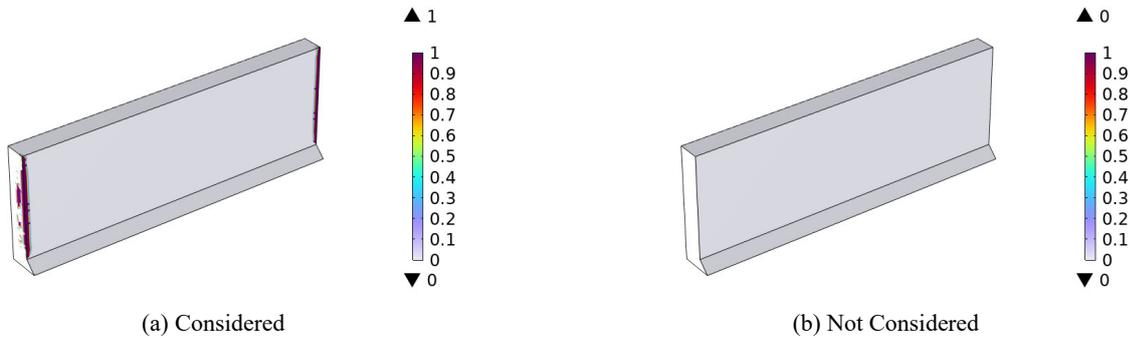

(a) Considered    (b) Not Considered

**Figure 4-15 Longitudinal dimension 15m: distribution of excess tensile strength**

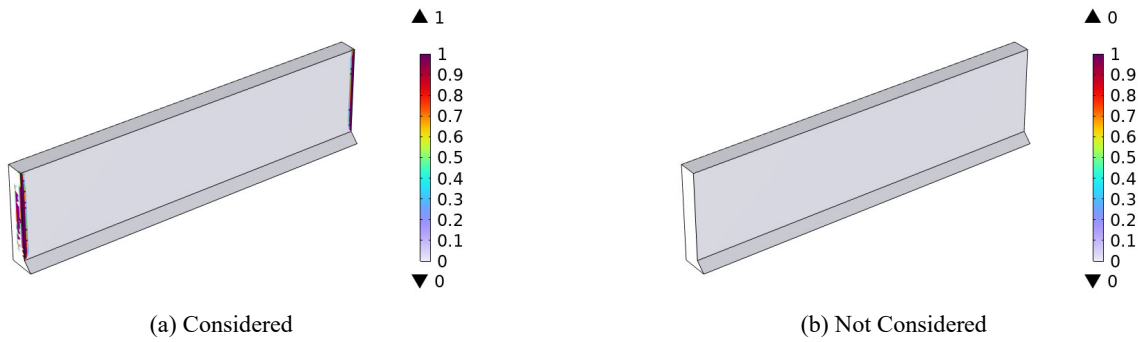

(a) Considered            (b) Not Considered

**Figure 4-16 Longitudinal dimension 20m: distribution of excess tensile strength**

Looking at the figure above, it can be seen that when considering the average forming temperature, regardless of how the longitudinal dimensions of the structure change, the tensile stress on both sides of the back wall of each size of concrete sidewall near the constrained area has exceeded its tensile strength standard value. According to the theory of maximum tensile stress, this situation indicates that the concrete will crack in this area, which may pose a serious threat to the safety of the structure. Without considering the part that does not exceed the standard value of tensile strength when the average forming temperature is taken into account, it proves that there is no possibility of cracking in this case.

Through comparative analysis, it can be obtained that if the strain reference temperature of the entire structure is set to the same constant value, that is, the initial temperature, that is, the average forming temperature is not taken into account, the concrete structure does not have the risk of cracking for the time being. However, when considering the average forming temperature, the tensile stress in the constrained edge area exceeds the tensile strength value. Comparing the two cases, it can be clearly seen that under low temperature conditions, the possibility of cracking in the mass concrete structure considering the average forming temperature is significantly increased compared with that without considering it, which also verifies the phenomenon mentioned above, which highlights the key role of the average forming temperature in evaluating the durability and stability of the mass concrete structure, and provides a strong basis for temperature control and crack prevention in subsequent engineering practice.

### ii. *Multidimensional comparison*

In order to explore in more depth whether to consider the average molding temperature and the specific impact on the temperature stress results, these two situations are analyzed from multiple dimensions。

（1） Structural overall stress comparison

First, for the overall level, the integral operation is carried out for the maximum tensile stress results obtained in the two cases, and the mean value is further obtained. The results are as follows:

From each model with different longitudinal dimensions, the first principal stress values of all nodes are extracted, and then systematically sorted out. The main function of the probability density distribution map is to describe the distribution characteristics of the data, so that the concentration trend, dispersion degree and distribution pattern of the stress distribution in the structure can be intuitively understood. Figures 4 13 to 4 16 show the probability density distribution maps of the first principal stress in the range of 1 MPa to 2.8 MPa and at intervals of 0.01 MPa under different longitudinal dimensions.

As can be seen from the above figure, when the longitudinal size is 5m, considering the average molding temperature, the maximum probability density is 218.06%/MPa corresponding to 2.71 MPa; when not considered, the maximum probability density is 691.03%/MPa corresponding to 1.44 MPa. When the longitudinal size is 10m, considering the average molding temperature, the maximum probability density is 295.37%/MPa corresponding to 2.76 MPa; when not considered, the maximum probability density is 672.88%/MPa corresponding to 1.72 MPa. When the longitudinal size is 15m, considering the average molding temperature, the maximum probability density is 314.45%/MPa corresponding to 2.78 MPa; when not considered, the maximum probability density is 728.95%/MPa corresponding to 1.45 MPa. When the longitudinal size is 20 meters, the maximum probability density is 345.66%/MPa corresponding to 2.78 MPa when considering the average molding temperature; when not considered, the maximum probability density is 828.11%/MPa corresponding to 1.45 MPa.

Comparing these figures, it can be found that when considering the average molding temperature, the probability density has a wide distribution range, almost all in the entire interval. When the average molding temperature is not considered, the probability density is only distributed in the range of 1.40MPa to 1.85MPa. With the increase of longitudinal size, when considering the average molding temperature, the probability density peaks are all 5, and the peak positions are relatively scattered, and the peak positions under different longitudinal sizes are similar; when the average molding temperature is not considered, the probability density peaks are about 6, and the positions are relatively dense, and the probability density values are much larger than the peaks when the average temperature is considered. It shows that when the average temperature is considered, the stress value distribution is wider; when the average molding temperature is not considered, the stress value distribution is more concentrated.

It is also observed from the figure that the stress probability density curves are similar under the same conditions for different longitudinal dimensions. In order to deeply compare the overall stress distribution in the two cases and between different longitudinal dimensions, the mean and standard deviation of the first principal stress distribution obtained in the two cases were obtained respectively. The results are shown in Table 4.1.

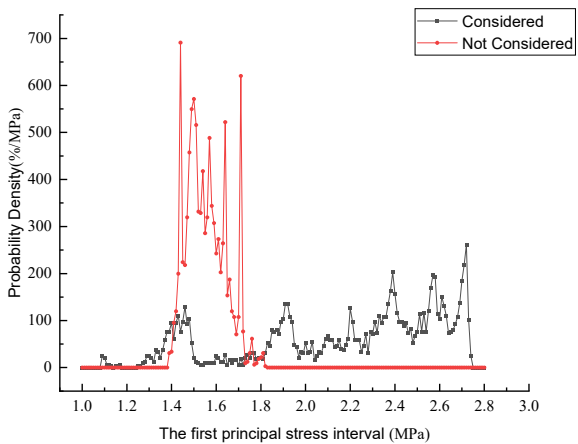

(a)5m

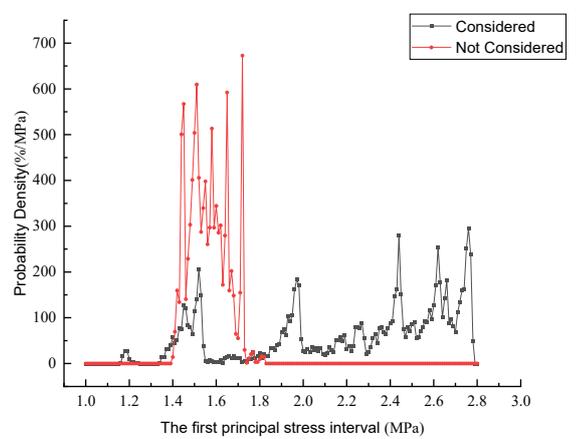

(b)10m

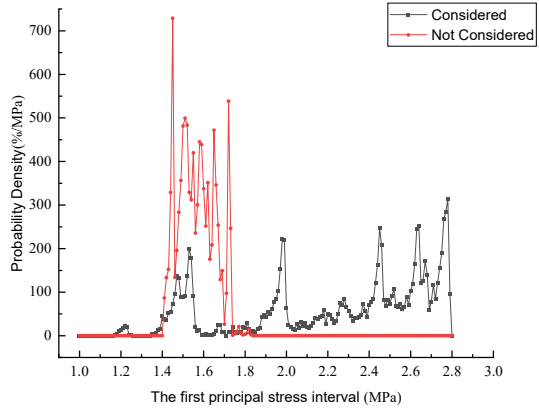
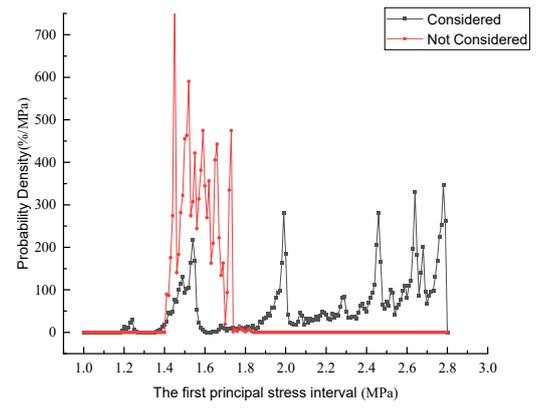

(c)15m  (d)20m

Table 4.1 Mean and standard deviation of the overall first principal stress

| Average molding temperature | Longitudinal dimension | | | | | | | |
|---|---|---|---|---|---|---|---|---|
| | 5m | | 10m | | 15m | | 20m | |
| | mean (MPa) | standard deviation $\sigma$ | mean (MPa) | standard deviation $\sigma$ | mean (MPa) | standard deviation $\sigma$ | mean (MPa) | standard deviation $\sigma$ |
| Consider Not | 2.29 | 0.30 | 2.30 | 0.30 | 2.31 | 0.29 | 2.32 | 0.29 |
| Considered | 1.56 | 0.12 | 1.57 | 0.13 | 1.57 | 0.13 | 1.57 | 0.13 |
| Difference (%) | 31.88 | - | 31.77 | - | 32.06 | - | 32.23 | - |

According to the data in the table, it is clear that under the same conditions, when the longitudinal size gradually increases, the overall mean stress changes very little. Among them, the maximum difference of the mean stress is only 0.03MPa when considering the average molding temperature; when the average molding temperature is not considered, the maximum difference is only 0.01MPa. At the same time, the standard deviation is almost the same. The maximum difference of the standard deviation is 0.01 in both cases. This fully shows that under the premise of other conditions being constant, the change of the longitudinal dimension has little effect on the distribution and magnitude of the temperature stress. Under the same longitudinal dimensions, the average first principal stress of the concrete as a whole when considering the average forming temperature is always higher than the value when this factor is not taken into account. The percentage difference between the two is about 32%, and the difference range is 0.73 to 0.75MPa, which is a relatively significant difference for concrete materials. This significant difference means that when the temperature history of the concrete forming period is fully taken into account and used as a critical condition for deformation, the overall temperature stress level caused by the low temperature shrinkage effect is significantly higher than that without fully considering the temperature factor during the forming period.

（2） Stress comparison of intermediate section

Then, the stress analysis is further carried out from the perspective of two-dimensional section, and it is also divided into whether to consider the average forming temperature. Take the longitudinal intermediate section of the model with different longitudinal dimensions, and derive the distribution of the first principal stress result. As shown in Figure 4 17 to Figure 4 20.

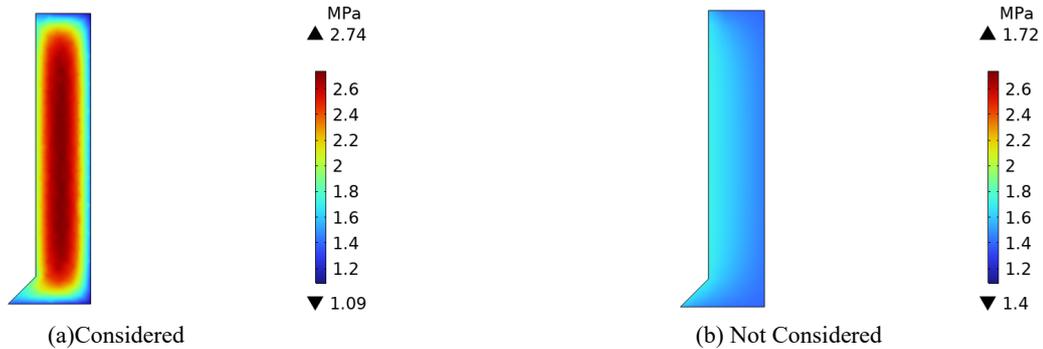

(a)Considered  (b) Not Considered
**Figure 4-17 Longitudinal dimension 5m: Comparison of maximum tensile stress distribution in intermediate section**

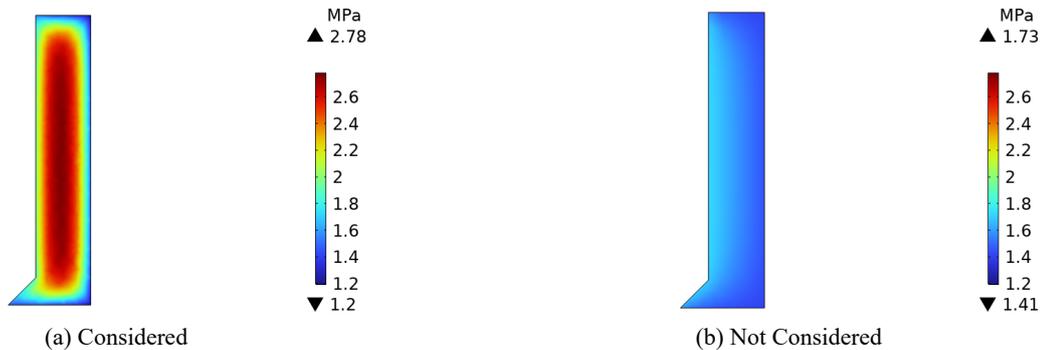

(a) Considered  (b) Not Considered
**Figure 4-18 Longitudinal dimension 10m: Comparison of maximum tensile stress distribution in intermediate section**

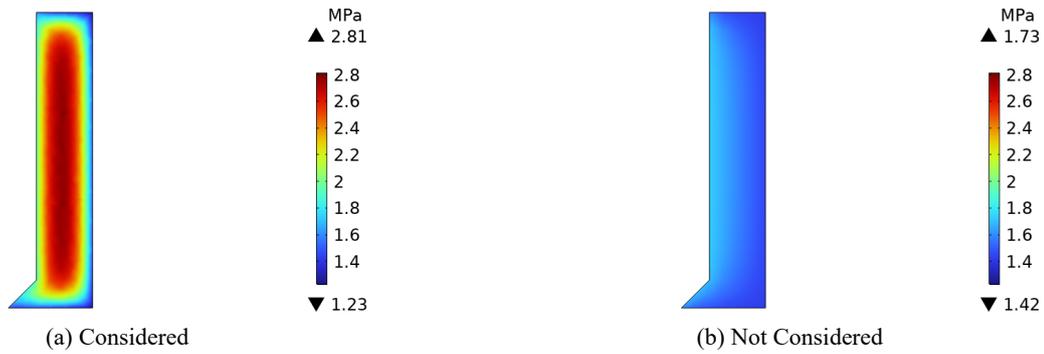

(a) Considered  (b) Not Considered
**Figure 4-19 Longitudinal dimension 15m: Comparison of maximum tensile stress distribution in intermediate section**

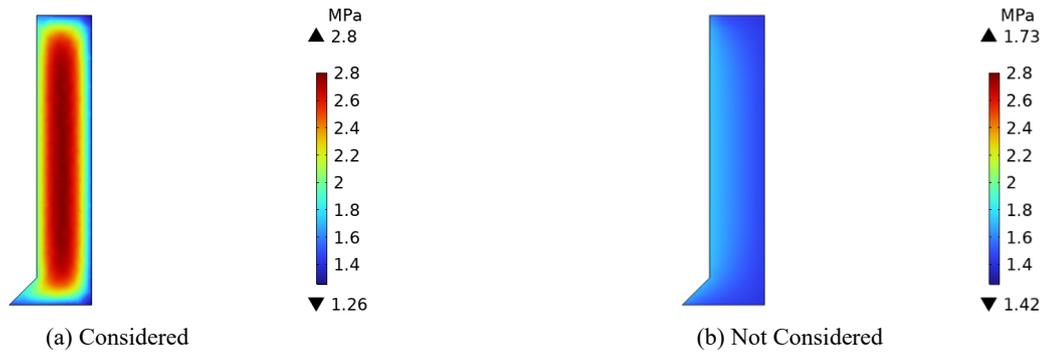

(a) Considered  (b) Not Considered
**Figure 4-20 Longitudinal dimension 20m: Comparison of maximum tensile stress distribution in intermediate section**

It can be seen from the figure that with the increase of longitudinal size, the stress range of each size is roughly 1.2 MPa~ 2.8 MPa. Without considering the average molding temperature, the stress range is about 1.4 MPa~ 1.73 MPa. It shows that under the same conditions, the stress distribution and range of the middle section of the concrete structure with different longitudinal sizes are almost the same.

From the analysis of the color legend, when considering the average forming temperature, the internal stress of the concrete structure section is significantly greater than the surface stress. This is because considering the average forming temperature is based on the temperature history during the forming period as a comprehensive consideration. Concrete is a material with poor thermal conductivity. For large-volume concrete, the internal heat is difficult to dissipate rapidly during the forming period, resulting in the internal temperature always higher than the surface temperature, which in turn makes the internal average forming temperature higher than the outer surface. When the average forming temperature is taken as the critical condition for temperature deformation, in the external low temperature environment, the internal deformation degree is greater than the outer surface, and when constrained, the internal tensile stress is greater. When the average forming temperature is not taken into account, the stress distribution shows a decreasing trend from the outside to the inside. This is because in this case, the critical condition of temperature deformation of the whole structure is set to the same constant value, while the outer surface of the concrete structure is in direct contact with the outside world. And the stress of the whole section is higher when considering the average forming temperature than when this factor is not taken into account.

In order to further confirm whether there is a difference in the first principal stress distribution of the intermediate section of the concrete structure with different longitudinal dimensions considering the average forming temperature and without considering the average forming temperature, the mean and standard deviation of the first principal stress distribution results of the intermediate section of the longitudinal dimension in each case are calculated. As shown in the table below。

**Table 4.2 Mean and standard deviation of the first principal stress in the intermediate section**

| average molding temperature | Longitudinal dimension | | | | | | | |
|---|---|---|---|---|---|---|---|---|
| | 5m | | 10m | | 15m | | 20m | |
| | Mean (MPa) | standard deviation $\sigma$ | mean (MPa) | standard deviation $\sigma$ | mean (MPa) | standard deviation $\sigma$ | mean (MPa) | standard deviation $\sigma$ |
| consider not | 2.29 | 0.43 | 2.30 | 0.42 | 2.31 | 0.41 | 2.31 | 0.41 |
| considered | 1.56 | 0.08 | 1.57 | 0.08 | 1.57 | 0.08 | 1.57 | 0.08 |
| difference (%) | 31.84 | - | 31.77 | - | 32.06 | - | 32.22 | - |

As can be seen from the above table, the mean stress of the intermediate section is almost equal with the change of longitudinal dimension regardless of the average forming temperature. Comparing the mean stress of the section at the same position with or without the average forming temperature, the mean stress considering the average forming temperature is about 0.73 to 0.74 MPa higher than that without considering it. This difference is similar to the comparison result of the overall average obtained before.

<span style="color:red">Stress contrast of central transection</span>

Since the temperature stress size and distribution are independent of the longitudinal size from the overall and two-dimensional cross-sectional analysis, the stress difference analysis in the two cases is now carried out from a one-dimensional perspective, and only the representative center section of the longitudinal size 5m middle section is selected, as shown in the figure below.

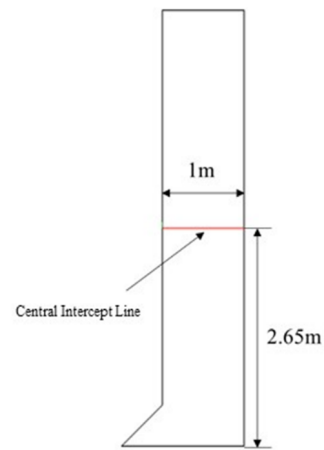

**Figure 4-21 Location of center section**

Extract the first principal stress value at each position of the cross-section, and compare the changing trend of the stress value at different positions in the two cases, as shown in Figure 4-22.

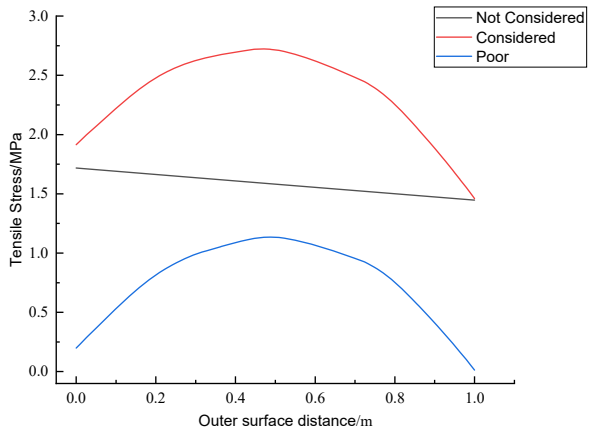
(a) 5m

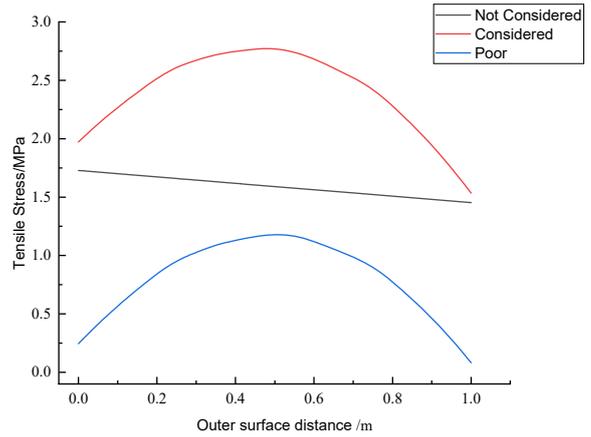
(b) 10m

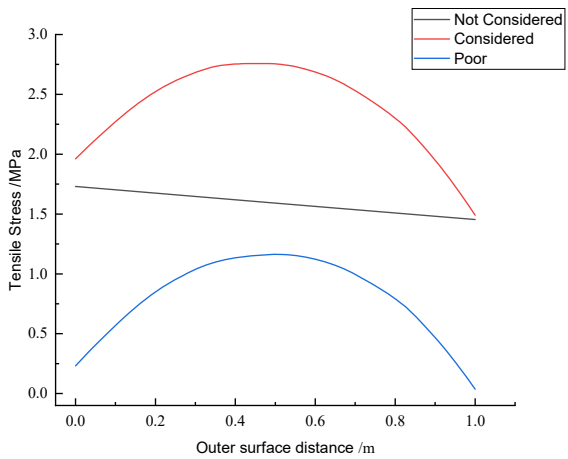
(c) 15m

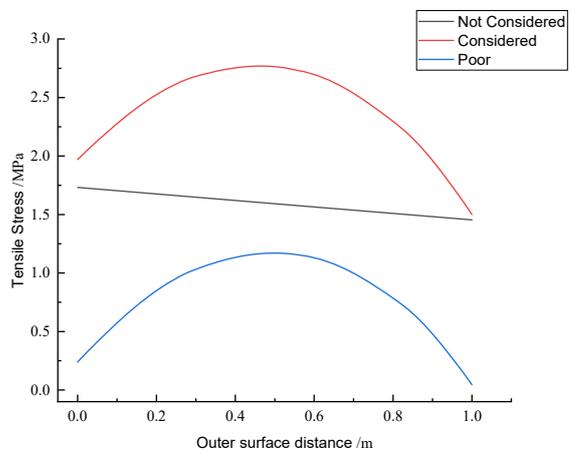
(d) 20m

**Fig. 4-22 Stress distribution at center section of different longitudinal dimensions**

The figure shows the variation trend of the first principal stress on the middle section with the increase of the distance from the outer surface and the difference of the tensile stress in the two cases with and without considering the average molding temperature. It can be seen from the figure that when considering the average molding temperature, the stress value first increases from 1.92 MPa to 2.74 MPa, and then gradually decreases to 1.51 MPa, showing a trend of increasing first and then decreasing, and reaching a peak at a distance of about 0.5 m from the outer surface. When the average molding temperature is not considered, the stress value decreases from 1.72 MPa to 1.45 MPa on the outer surface, showing a gradual decreasing trend, but the decrease ratio is not large. And the curve value when considering the average molding temperature is always greater than the value when it is not considered. The difference between the two cases also shows a trend of increasing first and then decreasing with the increase of the outer surface distance. The stress difference increases from 0.2 MPa to 1.15 MPa, and then decreases to the final 0.05 MPa. The maximum value of the difference also appears at a distance of about 0.5 meters from the outer surface.

（3） *Analysis of factor affecting temperature stress:*

Using the finite element model of the longitudinal dimension of the mass concrete side wall of the previous section of 5m, assuming that the concrete is fully formed, the factors affecting the temperature stress at 5 °C in winter environment are analyzed.

Different mold entry temperatures have a direct impact on the temperature rise inside the concrete, which in turn affects the average molding temperature results. The specification stipulates that the mold entry temperature of mass concrete should be controlled at 5 °C to 30 °C. In order to explore the influence of mold entry temperature on the average molding temperature and keep the remaining conditions unchanged, different concrete mold entry temperatures will be set in the temperature range of 5 °C to 30 °C according to a value of every 5 °C to solve the first principal stress distribution result at 5 °C in its cold environment.

The distribution of the first principal stress at different injection temperatures is shown in the figure below：

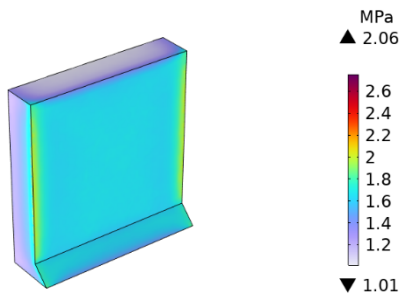
(a) Molding temperature 5℃

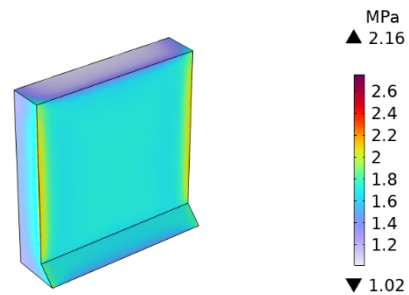
(b) Molding temperature 10℃

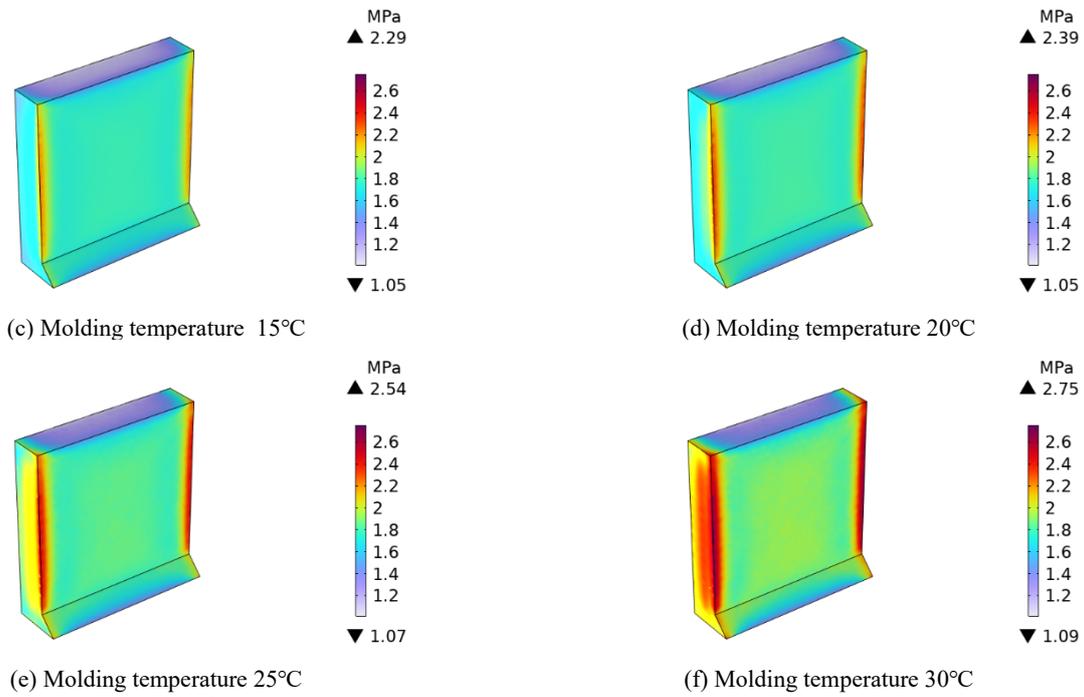

(c) Molding temperature 15°C  
(d) Molding temperature 20°C  
(e) Molding temperature 25°C  
(f) Molding temperature 30°C  

**Figure 4-23 Distribution of the first principal stress at different mold entry temperatures**

It can be seen from the figure that the stress at the inner center point gradually increases from 1.57MPa to 2.72MPa, and the growth rate also gradually increases, from 8.1% to 13.81%. The stress at the surface center point gradually increases from 1.67MPa to 1.92MPa, and the change of stress is roughly linear, and the change of growth rate is not significant. The stress at the boundary constraint point gradually increases from 2.00MPa to 2.73MPa, and the growth rate also gradually increases, from 5.13% to 8.37%.

The stress growth rate of the inner center point is the highest, which is due to the poor thermal conductivity of concrete, and the internal heat is difficult to dissipate. With the increase of the mold temperature, the temperature continues to rise, and the average molding temperature at this point also continues to rise, and the temperature stress generated at low temperatures is getting higher and higher. The surface center point is in direct contact with the external environment and is greatly affected by the external temperature. The average molding temperature growth rate is relatively flat, so the temperature stress growth is roughly linear. Although the boundary constraint point is also in direct contact with the outside world, this point is close to the constraint boundary and is more blocked during low temperature deformation. Although the stress change rate is not as high as that of the inner center point, the stress magnitude is always higher than that of other positions. The average overall stress gradually increases from 1.45MPa to 2.25MPa, and the growth rate also gradually increases, from 6.64% to 11.31%. It shows that with the increase of the temperature in the mold during pouring, after the concrete structure is fully formed, the overall temperature stress generated in the low temperature environment is gradually increasing.

（4） *Environmental Temperature Influence Analysis*

The ambient temperature during construction and pouring also has a direct impact on the temperature field of mass concrete, thus changing the result of the average forming temperature of concrete. In order to study the influence of ambient temperature on the average forming temperature, the mold entry temperature is fixed at 20 ° C, and the ambient temperature range from 10 ° C to 35 ° C is taken at intervals of 5 ° C, and each ambient temperature is ensured to be a constant value. On this basis, the average forming temperature of concrete under different ambient temperature conditions is calculated, and then the stress field at 5 ° C in winter is solved based on this.

Under the condition that the mold temperature remains the same, the distribution of the results of the first principal stress field corresponding to different ambient temperatures is shown in the figure below.

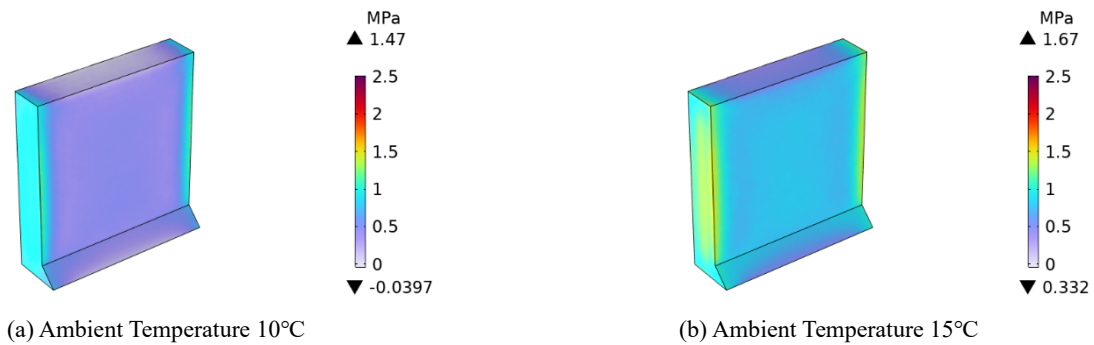

(a) Ambient Temperature 10°C  
(b) Ambient Temperature 15°C

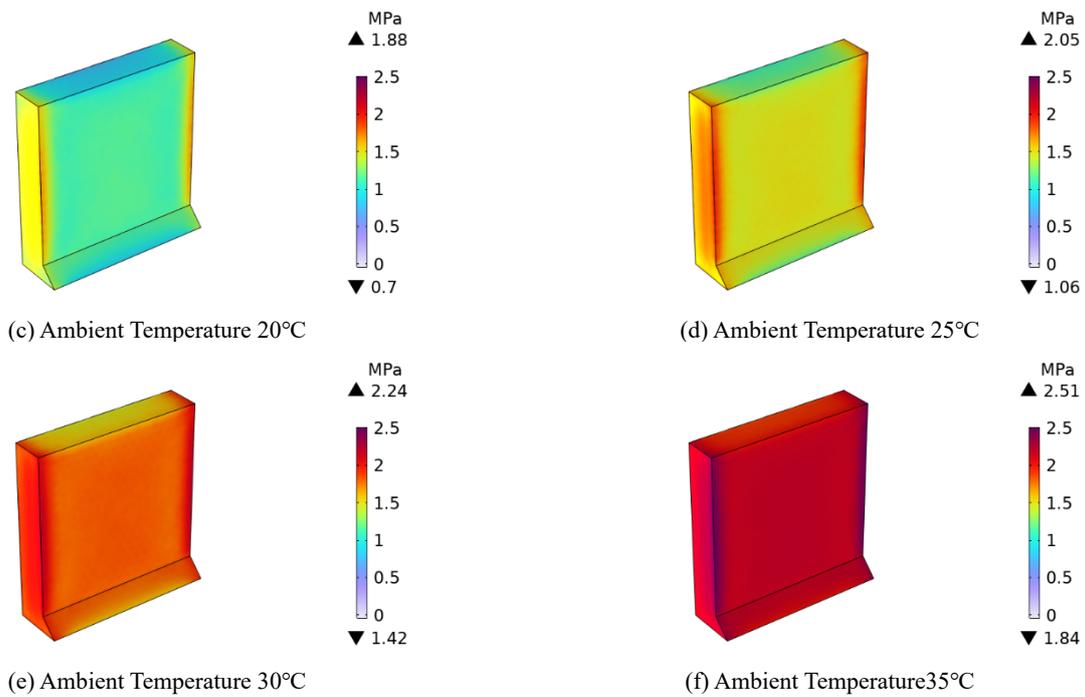

(c) Ambient Temperature 20°C  
(d) Ambient Temperature 25°C  
(e) Ambient Temperature 30°C  
(f) Ambient Temperature 35°C

**Figure 4-24 Average molding temperature distribution at different ambient temperatures**

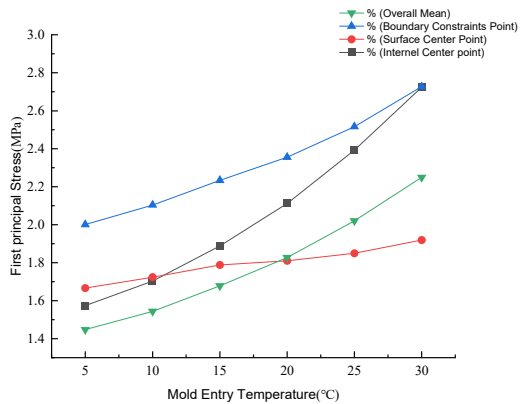

It can be seen from the figure that the stress growth at each point changes linearly, the inner center point stress gradually rises from 1.45MPa to 2.45MPa, and the average growth rate is 13.79%. The surface center point gradually rises from 0.45MPa to 2.19MPa, and the average growth rate is 77.33%. The boundary constraint point gradually rises from 1.20MPa to 2.50MPa, and the average growth rate is 21.67%.

The growth rate of the surface center point is the highest, which is due to the direct contact of the surface with the external environment, and the ambient temperature is set to a fixed value, and the surface is greatly affected by the external temperature. Although the boundary constraint point is also in direct contact with the external environment, the growth rate is not as high as the surface center point, and its stress value is always larger, indicating that the influence of the constraint conditions on the temperature stress is greater than that of the ambient temperature factor. As the ambient temperature increases, the stress difference between the internal center point and the surface point gradually decreases, indicating that the increase in ambient temperature decreases the temperature layer between the inside and the surface of the concrete, resulting in a decrease in the average molding temperature difference, resulting in a decrease in the temperature stress difference generated at the same low temperature. The average overall stress gradually increases from 0.93 MPa to 2.26 MPa, with an average growth rate of 28.60%, indicating that the ambient temperature has a positive impact on the overall temperature history of concrete, that is, the higher the temperature during the molding period, the greater the temperature stress generated in the low temperature environment after molding.

（5） Size Thickness Influence Analysis

The coagulation heat conductivity is poor and its thermal conductivity is low, which makes the internal heat conduction process of concrete relatively slow during the forming period. During the construction process, with the continuous increase of concrete pouring thickness, the accumulation of internal heat becomes more and more obvious. This is because the thicker concrete layer hinders the loss of heat to the external environment, making it difficult to release the internal hydration heat generated, which leads to the continuous increase of internal temperature. This heat accumulation and temperature increase caused by the change of pouring thickness has a significant impact on the temperature history of the concrete structure. Different pouring thicknesses will cause the concrete to show different temperature trends in different time periods, which in turn will also affect the average forming temperature.

In this study, the model is also constructed according to the longitudinal size of the tunnel concrete sidewall of 5 meters. The temperature of the mold entry is set at 20 °C, and the ambient temperature is constant at 20 °C. On the basis of the initial thickness, the thickness of the model is gradually changed by 0.5m each time, so as to further explore the influence of thickness changes on the temperature stress of concrete. The results of different thickness dimensions are shown in the figure below.

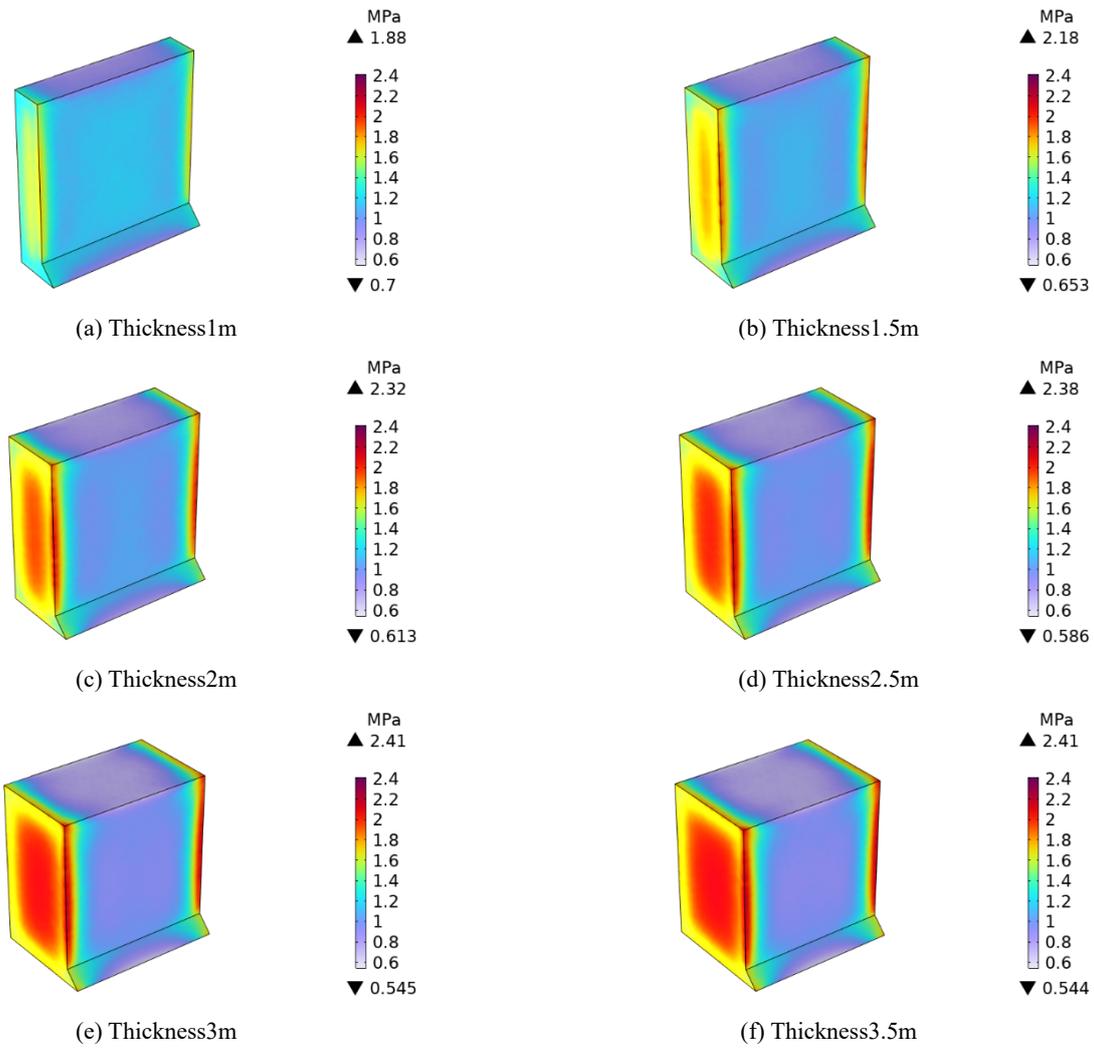

**Figure 4-25 Distribution of average molding temperature for different thicknesses**

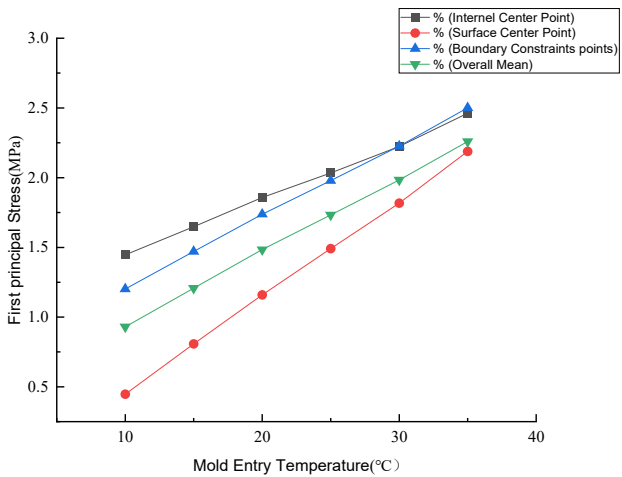

It can be seen from the figure that the stress at the inner center point gradually increases from 1.86MPa to 2.40MPa, and the growth rate is gradually decreasing, from 16.9% to 0.37%. The stress at the surface center point gradually increases from 0.90MPa to 1.16MPa, and the change of stress is very small. Although it increases linearly, the average growth rate is only 5.78%. The boundary constraint point stress gradually increases from 1.74MPa to 2.27MPa, and the growth rate is also gradually decreasing, from 12.78% to 1.21%.

As the thickness increases, the stress growth rate of the inner center point gradually becomes 0, mainly because the increase in the internal temperature of the concrete during the forming period mainly depends on the hydration and heat release of the cement, while the cooling mainly depends on the heat dissipation to the outside world. As the thickness of the concrete continues to increase, the internal center point gradually approaches the adiabatic state. Under this adiabatic trend, it becomes more and more difficult to dissipate heat outward, and the amount of heat dissipation continues to decrease. The temperature change is naturally not obvious, and the average molding temperature gradually does not change significantly.

Therefore, the temperature stress change generated under the same low temperature conditions after molding is very small. The stress value at the center point of the surface is smaller. This is due to the fact that the surface is in direct contact with the external environment, the heat dissipation is relatively fast, and the thickness change has no significant effect on the temperature of the surface point. Therefore, it has little effect on the average molding temperature, and the temperature stress generated under low temperature conditions is smaller. The growth trend of the boundary constraint point is similar to that of the internal center point. Although the boundary constraint point is also in contact with the external environment, its stress value is larger, which shows that the constraint condition has a significant impact on the size of the force. From the overall perspective, the average stress gradually increases from 1.48MPa to 1.90MPa, and the growth rate decreases from 12.78% to 1.21%. This indicates that the overall mean stress shows an upward trend with the increase of thickness. However, the growth rate gradually decreases due to the fact that most areas will gradually tend to an adiabatic state. At the same time, the overall mean stress is smaller than the inner center point stress because the lower stress on the surface part pulls down the overall mean.

## Discussion

The study evaluates the concept of average forming temperature in mass concrete and its impact on temperature-induced stress and cracking. The results indicate that incorporating average forming temperature in finite element simulations significantly influences the stress distribution within mass concrete structures. The study analyzed different longitudinal dimensions (5m, 10m, 15m, and 20m) and found that when the average forming temperature was considered, the tensile stress increased in the internal regions of the concrete. Figures 4-6 and 4-7 illustrate the modeling and meshing process for different concrete sidewalls, ensuring an accurate representation of the finite element model.

The temperature distribution over a 28-day hydration period was simulated, and the results demonstrated that the maximum tensile stress was higher when considering average forming temperature than when it was omitted. Figures 4-8 to 4-12 depict the temperature and stress distributions, emphasizing the impact of forming temperature history. The stress values presented in Table 4.2 confirm that neglecting the forming temperature underestimates the cracking risk. The stress comparison between sections, as shown in Figures 4-13 to 4-16, further reinforces this observation by highlighting areas where tensile stress exceeds concrete strength, posing a significant risk of cracking.

Additionally, Figure 4-22 presents the stress distribution at the center sections of various models, showing a peak stress value at around 0.5m from the outer surface. The comparison between considering and not considering the forming temperature reveals that the maximum difference in stress is approximately 1.17MPa, independent of the structure's size. This finding supports the argument that temperature-induced stress is not solely dependent on structural dimensions but is significantly affected by forming temperature history

## Conclusion

The study demonstrates that incorporating the concept of average forming temperature in mass concrete simulations enhances the accuracy of stress distribution predictions. The results suggest that considering forming temperature history can significantly impact temperature-induced stress and potential cracking risks. The maximum tensile stress, as shown in Table 4.2, confirms that the risk of cracking increases when forming temperature history is included, making it a critical factor in mass concrete design and construction.

Furthermore, the finite element analysis, validated through Figures 4-6 to 4-22, reveals that the tensile stress peaks at a certain distance from the surface, independent of structural size. This finding underscores the importance of accurate temperature field modeling in mass concrete structures. The methodology presented in this study offers a robust approach for predicting and mitigating thermal stress-induced cracking in large-scale infrastructure projects. Future research should explore advanced modeling techniques and material innovations to further optimize mass concrete durability and performance.

this study presents a robust methodology for improving the prediction of temperature stress and cracking risks in mass concrete. The inclusion of historical temperature effects is proven to be a critical factor in determining stress distribution, influencing structural durability. Future research should explore advanced hydration models and temperature-dependent material properties to further enhance crack prevention strategies in large-scale concrete infrastructure.

**Declaration of competing interest**
The authors declare that they have no known competing financial interests or personal relationships that could have appeared to influence the work reported in this paper.

**Data availability**
The data used can be obtained by contacting the corresponding author.

## References

[1] An, G., Yang, N., Li, Q., Hu, Y., & Yang, H. (2020). A Simplified Method for Real-Time Prediction of Temperature in Mass Concrete at Early Age. Applied Sciences, 10(13), 4451.

[2] Evsukoff, A.G., Fairbairn, E.M.R., Faria, É.F., Silvoso, M.M., & Toledo Filho, R.D. (2006). Modeling adiabatic temperature rise during concrete hydration: A data mining approach. Computers & Structures, 84(31-32), 2351-2362.

[3] Riding, K.A., Poole, J.L., Schindler, A.K., Juenger, M.C.G., & Folliard, K.J. (2006). Evaluation of Temperature Prediction Methods for Mass Concrete Members. ACI Materials Journal, 103(5), 357-365.

[4] Schackow, A., Effting, C., Gomes, I.R., Patruni, I.Z., Vicenzi, F., & Kramel, C. (2016). Temperature variation in concrete samples due to cement hydration. Applied Thermal Engineering, 103, 1362-1369.

[5] Wang, Z., Li, T., Yi, L., & Jiang, Y. (2018). Temperature Control Measures and Temperature Stress of Mass Concrete during Construction Period in High-Altitude Regions. Advances in Civil Engineering, 2018.

[6] Zhao, H., Jiang, K., Yang, R., Tang, Y., & Liu, J. (2020). Experimental and theoretical analysis on coupled effect of hydration, temperature and humidity in early-age cement-based materials. International Journal of Heat and Mass Transfer, 146, 118784.

[7] Bui, A. K., & Nguyen, T. C. (2020). The temperature field in mass concrete with different placing temperatures. *Civil Engineering and Architecture*, 8(2), 94-100.

[8] Ho, N. T., Nguyen, T. C., Bui, A. K., & Huynh, T. P. (2020). Temperature field in mass concrete at early-age: Experimental research and numerical simulation. *International Journal on Emerging Technologies*, 11(3), 936-941.

[9] Tia, M., Ferraro, C. C., Lawrence, A., Smith, S., & Ochiai, F. (2010). *Development of design parameters for mass concrete using finite element analysis: final report, February 2010* (No. UF Project No. 00054863). Florida Department of Transportation.

[10] Wang, L., Wang, Y., Miao, Y., Ju, S., Sui, S., Wang, F., ... & Jiang, J. (2024). Temperature damage assessment of mass concrete based on the coupling mechanism of hydration-temperature-humidity-constraint factors. *Journal of Building Engineering*, 90, 109211.

[11] Wei Guanhua, & Li Jian. (2019). Hydration heat research and simulation analysis of mass concrete. Sichuan Construction, (2), 309-314.

[12] Yan Peiyu, & Zheng Feng. (2006). Hydration kinetics model of cement-based materials. Journal of Silicate, 34 (5), 555-559.

[13] Scrivener, K., Ouzia, A., Juilland, P., & Mohamed, A. K. (2019). Advances in understanding cement hydration mechanisms. Cement and Concrete Research, 124, 105823.

[14] Ma Yuefeng. (2006). Study on Temperature and Stress of Concrete Based on Hydration Degree (Doctoral dissertation, Nanjing: Hohai University).

[15] Xue Hui, Liu Guangting, & Chen Fengqi. (1996). Development and application of a concrete adiabatic temperature rise tester. Journal of Tsinghua University: Natural Science Edition, 36 (1), 90-94.

[16] Hansen, P. F., & Pedersen, E. J. (1977). Maturity computer for controlled curing and hardening of concrete (No. Analytic).

[17] Gilliland, J. A. (2002). Thermal and shrinkage effects in high performance concrete structures during construction. The University of Calgary, Calgary, Albert, 9.

[18] Kaszyńska, M. (2002). Early age properties of high-strength/high-performance concrete. Cement and concrete composites, 24(2), 253-261.

[19] Poppe, A. M., & De Schutter, G. (2006). Analytical hydration model for filler rich binders in self-compacting concrete. Journal of Advanced Concrete Technology, 4(2), 259-266.

[20] Kim, S. G. (2010). Effect of heat generation from cement hydration on mass concrete placement.

[21] He, X. H., Cheng, H., & Qin, H. X. (2011). Study on temperature field and construction monitoring of hydration heat of railway cable-stayed bridge pile cap. Advanced Materials Research, 243, 1589-1596.

[22] Peng, J., Tang, C., Zhang, L., & Saeed, A. T. (2014). Experimental research on the hydration heat temperature field of hollow concrete piers. The Open Construction & Building Technology Journal, 8(1).

[23] Ismail, M., Noruzman, A. H., Bhutta, M. A. R., Yusuf, T. O., & Ogiri, I. H. (2016). Effect of vinyl acetate effluent in reducing heat of hydration of concrete. KSCE Journal of Civil Engineering, 20(1), 145-151.

[24] Mbogo, T. N. E. (2016). Portland Cement Paste Microstructure Characterization using Autogenous Shrinkage and Heat of Hydration. Asian Journal of Applied Science and Engineering, 5, 93-104.

[25] Lin, Y., & Chen, H. L. (2015). Thermal analysis and adiabatic calorimetry for early-age concrete members. Journal of Thermal Analysis and Calorimetry, 122, 937-945.

[26] Shanahan, N., Tran, V., & Zayed, A. (2017). Heat of hydration prediction for blended cements. Journal of Thermal Analysis and Calorimetry, 128, 1279-1291.

[27] Bažant, Z. P., Jirásek, M., Bažant, Z. P., & Jirásek, M. (2018). Temperature effect on water diffusion, hydration rate, creep and shrinkage. Creep and Hygrothermal Effects in Concrete Structures, 607-686.

[28] Klemczak, B., & Żmij, A. (2021). Insight into thermal stress distribution and required reinforcement reducing early-age cracking in mass foundation slabs. Materials, 14(3), 477.

[29] Chen, S. K., & Xu, B. W. (2021). Kinetic Model of Adiabatic Temperature Rise of Mass Concrete. Science of Advanced Materials, 13(5), 771-780.

[30] Liu, D., Zhang, W., Tang, Y., & Jian, Y. (2021). Prediction of hydration heat of mass concrete based on the SVR model. Ieee Access, 9, 62935-62945.

[31] Xie, Y., Du, W., Xu, Y., Peng, B., & Qian, C. (2023). Temperature field evolution of mass concrete: From hydration dynamics, finite element models to real concrete structure. Journal of Building Engineering, 65, 105699.

[32] Zhang Zengqi, Shi Mengxiao, Wang Qiang, & Cui Qiang. (2016). Accuracy of equivalent age method in performance prediction of mass concrete. Journal of Tsinghua University: Natural Science Edition, 56 (8), 806-810.

[33] Gaoyuan, & Zhang Jun. (2013). Prediction of early mechanical properties of concrete based on cement hydration degree. Engineering Mechanics, 30 (10), 133-139.

[34] Feng Chuqiao, Yu Xiaomin, Chang Xiaolin, Luo Daiming, & Xiong Jie. (2019). Derivation and application of kinetic model of chemical reaction of concrete hydration. China Rural Water Conservancy and Hydropower, (1), 152-157.

[35] Li Dong, & Zhang Yechen. (2021). Experimental analysis and calculation of concrete hydration exothermic model. Journal of Shanghai University/Shanghai Daxue Xuebao, 27 (4).